\documentclass[letterpaper, 10 pt, conference]{ieeeconf}  % Comment this line out if you need a4paper
\IEEEoverridecommandlockouts                              % This command is only needed if
\usepackage{graphicx} % for pdf, bitmapped graphics files
\usepackage{cite}
\usepackage{amsmath} % assumes amsmath package installed
\usepackage{amssymb}  % assumes amsmath package installed
\usepackage{braket}
\usepackage{booktabs}
\usepackage{algorithm}
\usepackage{algpseudocode}

\usepackage[caption=false, font=footnotesize]{subfig}

\usepackage{lipsum}

\newcommand{\mbx}{\mathbf{x}}
\newcommand{\mbu}{\mathbf{u}}
\newcommand{\mbm}{{\boldsymbol \mu}}
\newcommand{\mbe}{{\boldsymbol \epsilon}}
\newcommand{\argmin}{\mathop{\mathrm{argmin}}\limits}
\newcommand{\mcsymbol}[2]{\mathcal{#1}_{#2}}

\title{\LARGE \textbf{
Acceleration of Gradient-based Path Integral Method for \\
Efficient Optimal and Inverse Optimal Control
}}

\author{Masashi Okada$^{1}$ and Tadahiro Taniguchi$^{1,2}$% <-this % stops a space
\thanks{$^{1}$ AI Solutions Center, Business Innovation Division, Panasonic Corporation
        \texttt{\small okada.masashi001@jp.panasonic.com}}%
\thanks{$^{2}$ Ritsumeikan University, College of Information Science and Engineering
        \texttt{\small taniguchi@em.ci.ritsumei.ac.jp}}%
}

\begin{document}

\maketitle
\thispagestyle{empty}
\pagestyle{empty}

%%%%%%%%%%%%%%%%%%%%%%%%%%%%%%%%%%%%%%%%%%%%%%%%%%%%%%%%%%%%%%%%%%%%%%%%%%%%%%%%
\begin{abstract}
This paper deals with a new accelerated path integral method, which iteratively searches optimal controls with a small number of iterations.
This study is based on the recent observations that a path integral method for reinforcement learning can be interpreted as gradient descent.
This observation also applies to an iterative path integral method for optimal control, which sets a convincing argument for utilizing various optimization methods for gradient descent, such as momentum-based acceleration, step-size adaptation and their combination.
We introduce these types of methods to the path integral and demonstrate that momentum-based methods, like Nesterov Accelerated Gradient and Adam, can significantly improve the convergence rate to search for optimal controls in simulated control systems.
We also demonstrate that the accelerated path integral could improve the performance on model predictive control for various vehicle navigation tasks.
Finally, we represent this accelerated path integral method as a recurrent network, which is the accelerated version of the previously proposed path integral networks (PI-Net). We can train the accelerated PI-Net more efficiently for inverse optimal control with less RAM than the original PI-Net.
\end{abstract}

%%%%%%%%%%%%%%%%%%%%%%%%%%%%%%%%%%%%%%%%%%%%%%%%%%%%%%%%%%%%%%%%%%%%%%%%%%%%%%%%
\section{Introduction}
In recent years, research on the path integral control framework, which originated from Kappen's work \cite{kappen2005linear},
has significantly progressed.
One of the most remarkable works in the area is Williams's iterative path integral method \cite{williams2016aggressive, williams2017model}. This method iteratively updates a control sequence to search for the optimal solution on the basis of importance sampling of trajectories.
This approach solves the intrinsic problem of primitive path integral methods that require almost infinite samples for optimal solutions.
Moreover, Williams derived different iterative methods \cite{Williams-ICRA-17, williams2017information}, which eliminate the affine dynamics constraints on the original path integral framework.
This paper focuses on this type of iterative path integral methods.

Since path integral methods are derivative free, they possess several attractive features.
First, as it is not necessary to approximate dynamics and cost models with linear and quadratic forms,
non linear system dynamics and cost functions can be naturally employed.
In~\cite{williams2016aggressive, williams2017information}, highly non linear car dynamics have been employed to successfully control a miniature real vehicle that functions autonomously and with aggressive drifting.
Dynamics can also be represented as trainable models, i.e., neural networks, thus allowing to solve model-based reinforcement learning tasks.
Aggressive driving was also performed using the model-based reinforcement learning of neural dynamics~\cite{Williams-ICRA-17}.

Trainable cost models could be introduced to path integral methods;
the problem would be how to train the parameterized cost models. This is known as inverse reinforcement learning or inverse optimal control problem.
Okada proposed a solution introducing path integral networks (PI-Net) \cite{okada2017path} which are recurrent networks to completely imitate the iterative path integral method.
This network is differentiable with the parameters in the network and thus the network can be trained by back-propagation in the same way as done for standard neural networks.
Inverse optimal control can be simply conducted by training PI-Net via imitation learning.

It has been claimed \cite{williams2017model} that the convergence rate of the iterative method is fast enough to apply it to model predictive control (MPC \cite{camacho2013model}; a.k.a. receding horizon control), however, the convergence performance has not been clearly determined.
On the other hand, the experiments of PI-Net in \cite{okada2017path} required hundreds of iterations (or network recurrences) to obtain good training results, leading to massive RAM usage during back-propagation.

Based on the above, this paper discusses the convergence of iterative path integral methods, using the observations from recent work by Miyashita et al.~\cite{miyashita2017mirror}.
In that report, it is shown that a path integral method for reinforcement learning (PI${}^2$, Policy Improvement with Path Integral \cite{theodorou2010generalized}) can be derived from a variant of a gradient descent method (i.e., mirror descent \cite{bubeck2015convex}), showing a connection between PI${}^2$ and gradient descent.
Accordingly, a connection between the iterative path integral method and gradient descent can be expected.
In fact, as pointed out later in Sect.~\ref{sec:derivation}, such connection can be derived using the same concepts as \cite{miyashita2017mirror}.
Considering the iterative path integral as gradient descent, this paper aims to accelerate the convergence of the iterative method utilizing optimization methods that were developed to accelerate the gradient descent.

The organization and contributions of this paper are summarized as follows.
1) Sect.~\ref{sec:derivation} clarifies the relation between the iterative path integral method and gradient descent.
It is shown that a new iterative path integral method is derived from the mirror descent search \cite{miyashita2017mirror}.
% We remark that this derivation strongly owes to the previous work of \cite{miyashita2017mirror}.
%
2) In Sect.~\ref{sec:accel}, we introduce optimization methods for gradient descent to accelerate the iterative method.  
We also discuss how the accelerated method is applied to MPC and PI-Net.
3) Finally, in Sect.~\ref{sec:exp}, we conduct simulated experiments of four dynamics systems. 
The experiment shows that the accelerated methods could dramatically accelerate the convergence compared to the original method.
It is also shown that MPC and inverse optimal control benefit from the acceleration method.

\section{Preliminaries}
This section briefly reviews the formulation of stochastic optimal control problem, Williams's iterative path integral methods \cite{williams2016aggressive,williams2017model,williams2017information,Williams-ICRA-17}, and PI-Net \cite{okada2017path}.

\subsection{Formulation of Stochastic Optimal Control}
In this paper a discrete-time, continuous dynamics taking the following form is assumed:
\begin{equation}
	\mbx_{t+1} = f(\mbx_{t}, \mbu_{t}),
	\label{eqn:dynamics}
\end{equation}
where $\mbx_{t} \in \mathbb{R}^{n}$ is the state of the system at time $t$, $\mbu_{t} \in \mathbb{R}^{m}$ is the control input at time $t$, and $f: \mathbb{R}^{n}\rightarrow \mathbb{R}^{n}$ denotes the state-transition function of the system.
We assume that $\mbu_{t}$ is a stochastic variable that takes the form:
\begin{equation}
	\mbu_{t} = \mbm_{t} + \mbe_{t}, \label{eqn:ctrl_w_noise}
\end{equation}
where $\mbm_{t} \in \mathbb{R}^{m}$ is the deterministic variable and $\mbe_{t} \in \mathbb{R}^{m}$ is the stochastic variable which represents the system noise with zero mean, namely $\mathbb{E}[\mbu_{t}] = \mbm_{t}$.
Given a finite time-horizon $t \in \{0, 1, 2, \cdots, T - 1\}$, state-action trajectory $\tau = \{\mbx_{0}, \mbu_{0}, \mbx_{1}, \cdots, \mbu_{T-1}, \mbx_{T}\}$ by (\ref{eqn:dynamics}) and trajectory cost $S(\tau) \in \mathbb{R}$, the objective of the stochastic optimal control problem is to find a control sequence $\mbm_{0:T-1}^{*} = (\mbm^{*}_{0}, \mbm^{*}_{1}, \cdots, \mbm^{*}_{T-1}) \in \mathbb{R}^{m \times T}$ which minimize the objective function $J = \mathbb{E}\left[S(\tau)\right]$.
The optimal problem is formulated as:
\begin{equation}
  \mbm_{0:T-1}^{*} = \argmin J = \argmin\mathbb{E}\left[S(\tau)\right]. \label{eqn:soc_form}
\end{equation}
In the following discussion, $\mbm_{0:T-1}$ is denoted as $\mbm$ for readability.
We used same notation for $\mbu \equiv \mbu_{0:T-1} = (\mbu_{0}, \mbu_{1}, \cdots, \mbu_{T-1})$ and $\mbe \equiv \mbe_{0:T-1} = (\mbe_{0}, \mbe_{1}, \cdots, \mbe_{T-1})$.

\subsection{Iterative Path Integral Methods} \label{ssec:iterative}
The iterative path integral methods \cite{williams2016aggressive,williams2017model,williams2017information,Williams-ICRA-17} are optimization methods for the stochastic optimal control problem.
These methods assume that the system noise $\mbe_{t}$ is zero-mean Gaussian $\mbe_{t} \sim \mathcal{N}(\mathbf{0}, \Sigma)$ with a covariance matrix $\Sigma \in \mathbb{R}^{m \times m}$, and suppose a trajectory cost function $S(\tau)$ as the sum of arbitrary state-cost and quadratic control-cost over time time-horizon:
\begin{equation}
	\begin{split}
	S(\tau) &= S_{\mbx}(\tau) + S_{\mbu}(\tau), \\
    S_{\mbx}(\tau) &= \phi(\mbx_{T}) + \sum^{T-1}_{t=0}q(\mbx_{t}), \\
    S_{\mbu}(\tau) &= \sum^{T-1}_{t=0}\mbu^{\mathrm{T}}_{t}R\mbu_{t},
    \label{eqn:traj_cost}
    \end{split}
\end{equation}
where $\phi: \mathbb{R}^n \rightarrow \mathbb{R}$ and $q: \mathbb{R}^n \rightarrow \mathbb{R}$ are respectively terminal- and running-cost, and $R \in \mathbb{R}^{m\times m}$ is a weight matrix for the quadratic control cost.

These methods solve the optimal problem by iteratively updating $\mbm$ with the equation:
\begin{equation}
  \begin{split}
    \mbm^{(j)} &= \mbm^{(j-1)} + \Delta\mbm^{(j-1)}, \\
    \Delta\mbm^{(j-1)} &= \frac{\mathbb{E}_{p^{(j-1)}}\left[e^{-\tilde{S}(\tau)/\lambda}\mbe\right]}{\mathbb{E}_{p^{(j-1)}}\left[e^{-\tilde{S}(\tau)/\lambda}\right]},
  \end{split}\label{eqn:update_law}
\end{equation}
where $j$ is the update iteration index, $p^{(j-1)}$ is the probability density function of the stochastic variable $\mbu$, $\lambda \in \mathbb{R}^{+}$ is a hyper-parameter called the inverse temperature, and $\tilde{S}(\tau)$ is the modified trajectory cost function. For example, Williams et al.~\cite{williams2016aggressive} define $\tilde{S}(\tau)$ as:
\begin{equation}
	\tilde{S}(\tau) = S_{\mbx}(\tau) + \sum^{T-1}_{i=0}\mbm^{\mathrm{T}}_{t} R \mbe_{t}. \label{eqn:S_tilde}
\end{equation}
% \begin{equation}
% 	p^{(j-1)} = \mathcal{N}\left(\mbm^{(j-1)}, \Sigma\right).
% \end{equation}
Although different concepts are adopted to derive the methods in \cite{williams2017model} and \cite{Williams-ICRA-17,williams2017information} %
\footnote{
Ref.~\cite{williams2017model} is based on the linearization of the Hamilton-Jacob Bellman equation and application of Feyman-Kac lemma, whereas \cite{Williams-ICRA-17,williams2017information} adopt information theoretic approach using KL divergence and free energy.
}%
, they are practically equivalent and theoretically related.
Specifically the method in \cite{Williams-ICRA-17} and \cite{williams2017information} can exactly recover the method in \cite{williams2017model} if dynamics is constrained to be affine in control.
% %
% \footnote{
% PI-OC derivation assumes control affine dynamics
% whereas IT-OC can represent dynamics by arbitrary non-linear functions
% }.
%
Eq.~(\ref{eqn:update_law}) can be implemented on digital computers
by approximating the expectation value with the Monte Carlo integral as shown in Alg.~\ref{alg:pi},
where $K$ is the number of trajectories to be simulated for the approximation.
%%%% Pseudo Code of SOC %%%%
\begin{algorithm}[tbh]
\small
\caption{\small Iterative Path Integral Method \cite{williams2016aggressive,williams2017model,williams2017information,Williams-ICRA-17}} \label{alg:pi}
\begin{algorithmic}[1]
% \INPUT $K$, $T$: Number of trajectories \& timesteps \\
\Require $\mbx_{0}$, $\mbm^{(j-1)}$ : Input state \& Control sequence
% $\mbx_{0}$ : Input State \\
% $\mbm^{(j-1)}$: Control sequence\\
% $f$, $S$: Dynamics and trajectory cost function \\
% $\lambda$: Hyper-parameter
\Ensure $\mbm^{(j)}$: Improved control sequence 
%
% Initialization
\State Sample $\{\mbe^{(0)}, \mbe^{(1)}, \cdots, \mbe^{(K-1)}\}$
% \STATE $\mathcal{T} \leftarrow \emptyset$
%
% Monte-Carlo simulation
\For{$k \leftarrow 0$ \textbf{to} $K-1$}
\State $\mbx^{(k)}_{0} \leftarrow \mbx_{0}$
\State $\tau^{(k)} \leftarrow \emptyset$
\For{$t \leftarrow 0$ \textbf{to} $T-1$}
\State $\tau^{(k)} \leftarrow \tau^{(k)} \cap \{\mbx^{(k)}_{t}, \mbu^{(k)}_{t}\}$
\State $\mbu^{(k)}_{t} \leftarrow \mbm^{(j-1)}_{t} + \mbe^{(k)}_{t}$
\State $\mbx^{(k)}_{t + 1} \leftarrow f(\mbx^{(k)}_{t}, \mbu^{(k)}_{t})$
\EndFor
\State $\tau^{(k)} \leftarrow \tau^{(k)} \cap \{\mbx^{(k)}_{T}\}$
% \STATE $\mathcal{T} \leftarrow \mathcal{T} \cap \{\tau^{(k)}\}$
\EndFor
%
% Update control sequence
\State $\Delta\mbm^{(j-1)} \leftarrow {\sum^{K-1}_{k=0}e^{-\tilde{S}(\tau^{(k)}) / \lambda}\mbe^{(k)}} \big/ {\sum^{K-1}_{k=0}e^{-\tilde{S}(\tau^{(k)}) / \lambda}}$
% \STATE Calculate $\Delta\mbm^{(j-1)}$ with $\mathcal{T}$, $\mathcal{E}$ and Eq.~(\ref{eqn:mc_approx}).
% \STATE $\Delta\mbm \leftarrow \textsc{CalculateStep}\left(S, \mathcal{E}, \mathcal{T}\right)$
\State $\mbm^{(j)} \leftarrow \mbm^{(j-1)} + \Delta\mbm^{(j-1)}$
\end{algorithmic}
\end{algorithm}

%%%%%%%%%%%%%%%%%%%%%%%%%%%%
%
% Different from other general optimal control algorithms, such as Differentiable Dynamics Programming (DDP) \cite{todorov2005generalized}, SOC does not require first or second-order approximation of the dynamics and a quadratic approximation of the cost model, naturally allowing for non-linear system dynamics and cost models.
% For instance, experiments in \cite{williams2017model} demonstrate the PI-OC outperforms DDP on model predictive control tasks with the non-linear dynamics and cost models.
% In addition, this flexibility allows us to use general function approximators, such as neural networks, to represent dynamics.
% Thus SOC is very compatible with model-based reinforcement learning \cite{Williams-ICRA-17}.

Let $U$ be the iteration numbers $(j \in \{0, 1, \cdots, U-1\})$, the computational complexity of the iterative methods is $O(U \times T \times K)$.
Since we can independently simulate $K$ trajectories ($\ell$2--11 in Alg.~\ref{alg:pi}),
GPU parallelization is feasible and would significantly reduce the effect of $K$, achieving real-time MPC~\cite{williams2017model}.
% Previous works \cite{williams2016aggressive, williams2017model, Williams-ICRA-17, williams2017information} have used $U = 1$ for MPC.
% however, it has not been discussed how $U$ effects to optimization results.
% Knowing the effect $U$ and other SOC parameters will be necessary to establish the design methodology for practical systems.

\subsection{Path Integral Networks}
Okada et al.~\cite{okada2017path} regarded the iterative executions of Alg.~\ref{alg:pi} as a \textit{double-looped} recurrent network, termed PI-Net%
\footnote{
Please see Fig.~\ref{fig:pi_arch} for an overview of the recurrent architecture of PI-Net.
Although the figure shows the architecture of accelerated PI-Net introduced in Sect.~\ref{sec:accel}, the difference is slight and removing the flows of $\Delta\mbm$ (also introduced later) makes it completely equivalent to the original.
}.
%
%
% the outer loop for the iteration of Alg.~\ref{alg:pi} and the inner loop for the sequential state predictions in $\ell$5--9.
The dynamics $f$ and/or cost models $S(\tau)$ in the network can be represented as trainable models such as neural networks.
Since the network is fully differentiable, we can optimize the internal model parameters by end-to-end training of PI-Net by back-propagation.
This property of the network can be utilized for inverse optimal control to learn cost models latent in experts' demonstrations.
Inverse optimal control with PI-Net is simply achieved by imitation learning, in which PI-Net is supervisedly trained so that the network output mimics the experts' demonstrations.

Let $B$ be the mini-batch size for PI-Net training.
Then the back-propagation requires a RAM size of $O(B \times U \times T \times K)$, which immediately grows, forcing to use massive of RAM.
For instance in \cite{okada2017path}, hundreds of giga-bytes RAM were used even though they focused on rather simple tasks.
Therefore, schemes to reduce the parameters while achieving good training results are necessary.

\section{Connection of Iterative Path Integral to Gradient Descent} \label{sec:derivation}
This section shows the connection between iterative path integral methods and gradient descent.
Relations between the iterative methods are also discussed.

\subsection{Iterative Path Integral Method from Mirror Descent}
Miyashita et al.~employed the mirror descent \cite{miyashita2017mirror} to policy search of parameterized model so as to maximize the expected cumulative reward, deriving PI${}^2$.
Since the derivation approach is very general, it is easily applied to optimal control search by regarding control sequence $\mbm$ as policy parameters.
The following brief derivation results in a different iterative path integral method.
Details of the derivation concept can be found in \cite{miyashita2017mirror}, as the present procedure is based on that works.
% The derivation is highly influenced by , so readers who are interested in the details are recommended to see the reference.

We consider to optimize $p^{(j-1)}$ in order to minimize the objective function $J$ by iteratively updating $p^{(j-1)}$ using the mirror descent:
\begin{equation}
	p^{(j)} = \argmin_{p} \braket{g, p} + \frac{1}{\alpha} D_{\mathrm{KL}}(p, p^{(j-1)}) + \beta\left(1 - \int dP \right),
    \label{eqn:mgd}
\end{equation}
where $g$ is the gradient of the objective $J$ with respect to $p$, $\braket{\cdot, \cdot}$ is the cross-product operator and $\alpha$ is a hyper-parameter corresponding to the step-size. $\beta$ is the Lagrange multiplier for the constraint $\int dP = 1$ where $P$ is the cumulative distribution of $p$.
$D_{\mathrm{KL}}(p, p^{(j-1)})$ is the KL divergence between two probability density function defined as:
\begin{equation}
	D_{\mathrm{KL}}(p, p^{(j-1)}) = \int \log\frac{p}{p^{(j-1)}} dP.
\end{equation}
The gradient $g$ can be calculated from:
\begin{equation}
	g = \frac{\partial J}{\partial p} = \frac{1}{\partial p}\int p S(\tau) d\mbu = S(\tau)d\mbu,
\end{equation}
then the inner-product term becomes:
\begin{equation}
	\braket{g, p} = \int S(\tau) p d\mbu = \int S(\tau)dP.
\end{equation}
Using the above relations, we can represent the argument of $\argmin$ in (\ref{eqn:mgd}) as:
\begin{equation}
	\int\left(
    	S(\tau) + \frac{1}{\alpha}\log\frac{p}{p^{(j-1)}} - \beta \right) dP + \beta.
\end{equation}
Organizing the above equation yields:
\begin{equation}
    \frac{1}{\alpha} D_{\mathrm{KL}}\left(p, p^{(j-1)} e^{\alpha(\beta - S(\tau))}\right) + \beta.
\end{equation}
By minimizing this equation, the update law with respect to $p$ is derived as:
\begin{equation}
	p^{(j)} = p^{(j-1)}e^{\alpha(\beta - S(\tau))}.
\end{equation}
The Lagrange multiplier $\beta$ can be removed using the constraint $\int dP^{(j)} = 1$, then:
\begin{equation}
	p^{(j)} = \frac{p^{(j-1)}e^{-\alpha S(\tau)}}{\int e^{-\alpha S(\tau)} dP^{(j-1)}}
    = \frac{p^{(j-1)}e^{-\alpha S(\tau)}}{\mathbb{E}_{p^{(j-1)}}\left[e^{-\alpha S(\tau)}\right]}. \label{eqn:update_law_p}
\end{equation}

Next, we derive the update law for $\mbm$ from (\ref{eqn:update_law_p}) with the relation $\int \mbu dP^{(j)} = \mbm^{(j)}$.
By multiplying both sides of (\ref{eqn:update_law_p}) by $\mbu d\mbu$ and integrating both sides:
\begin{equation}
 	\mbm^{(j)} = \frac{\int e^{-\alpha S(\tau)}\mbu dP^{(j-1)}}{\mathbb{E}_{p^{(j-1)}}\left[ e^{- \alpha S(\tau)}\right]}. \label{eqn:update_law_immature}
\end{equation}
Substituting (\ref{eqn:ctrl_w_noise}) in (\ref{eqn:update_law_immature}) leads to:
\begin{equation}
 	\mbm^{(j)} = \mbm^{(j-1)} + \frac{\int e^{- \alpha S(\tau)}\mbe dP^{(j-1)}}{\mathbb{E}_{p^{(j-1)}}\left[e^{- \alpha S(\tau)}\right]}, \label{eqn:update_law_mu0}
\end{equation}
\begin{equation}
	\therefore \mbm^{(j)} = \mbm^{(j-1)} + \frac{\mathbb{E}_{p^{(j-1)}}\left[e^{- S(\tau) / \lambda}\mbe\right]}{\mathbb{E}_{p^{(j-1)}}\left[e^{- S(\tau)/ \lambda}\right]}. \label{eqn:update_law_mu}
\end{equation}
At the final step, we replaced $\alpha$ with $1 / \lambda$ to emphasize the similarity between the original methods and the newly derived method.
If we replace $S(\tau)$ with $\tilde{S}(\tau)$, the derived and original methods are completely equivalent.

The convergence rate of the mirror descent is theoretically proved to be $O(1/j)$ \cite{krichene2015accelerated} and this would be valid in the the derived method.
Considering the similarity of methods, we also assume that the original method has the same convergence.

\subsection{Differences Between the Iterative Path Integral Methods}
As summarized in Appx.~\ref{app:it_derivation}, the original path integral method \cite{Williams-ICRA-17} is derived based on iterative importance sampling.
It is interesting to note that different concepts (i.e., importance sampling and mirror descent) result in similar iterative methods.
However, it may be said that gradient descent is a more adequate method for optimal solution search. 
In addition, the fact that the iterative method is gradient-based allows us to utilize a variety of optimization methods developed for gradient descent.
Furthermore, the derivation with mirror descent did not use several assumptions that are necessary for the original methods.
This is summarized in Table~\ref{tab:comp_soc}, which also shows that the newly derived method is a generalization of the original ones.
Since we mainly focused on the convergence performance, the generalized property will not be examined in this paper.
The experiments in Sect.~\ref{sec:exp} consider the same assumptions as in references \cite{Williams-ICRA-17,williams2017information}.
\begin{table}[t]
	\centering
    \caption{Assumptions on iterative path integral methods} \label{tab:comp_soc}
    \begin{tabular}{c|ccc}
    	\toprule
    	& \cite{williams2016aggressive,williams2017model} & \cite{Williams-ICRA-17,williams2017information} & Derived \\\hline
    	Dynamics & Control Affine & \textbf{Arbitrary} & \textbf{Arbitrary} \\
        Traj. Cost & Eq.~(\ref{eqn:traj_cost}) & Eq.~(\ref{eqn:traj_cost}) & \textbf{Arbitrary} \\
        Sys. Noise & Gaussian & Gaussian & \textbf{Arbitrary} \\
        \bottomrule
    \end{tabular}
\end{table}

\section{Employing Optimization Methods for Gradient Descent} \label{sec:accel}
This section introduces a variety of optimization methods for gradient descent in the iterative path integral method.
We also describe the application of one method to MPC and PI-Net.

\subsubsection{Nesterov Accelerated Gradient (NAG) \cite{nesterov1983method}}
\begin{figure}
	\centering
    \includegraphics[width=0.475\textwidth]{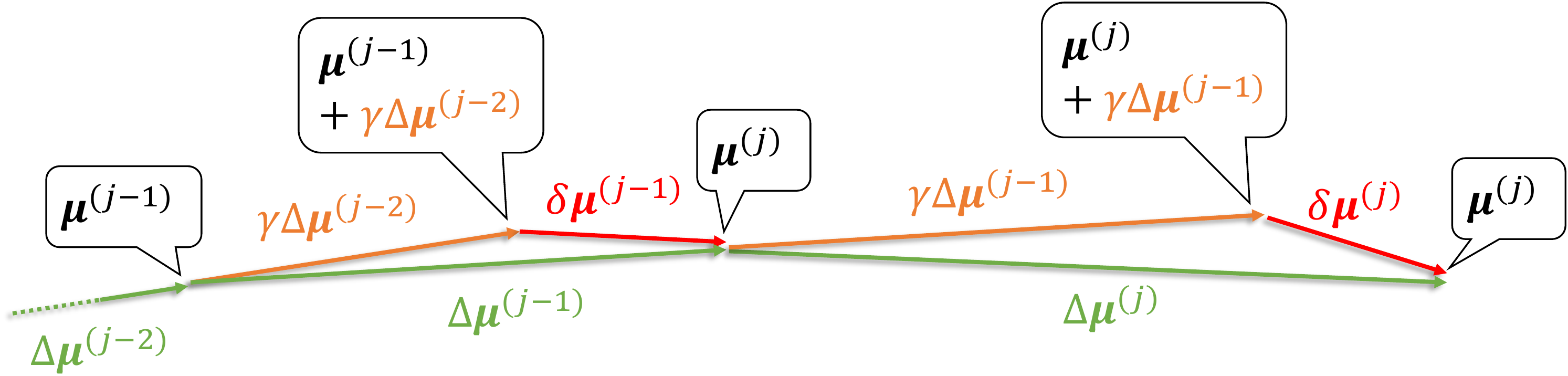}
    \caption{
    Nesterov Accelerated Gradient: accumulation of past momenta accelerate the convergence.
    } \label{fig:nag}
\end{figure}

NAG utilizes the past update (or momentum); in the present work $\Delta\mbm^{(j-2)} = \mbm^{(j-1)} - \mbm^{(j-2)}$. 
In this method: (1) a current solution is \textit{drifted} by the momentum and then (2) the drifted solution is \textit{slided} by the gradient.
The iterative path integral method can simply adopt this concept.
Let us consider a probability density function $p_{m}^{(j-1)}$ in which the mean value is drifted from $\mbm^{(j-1)}$ by the momentum $\Delta\mbm^{(j-2)}$:
\begin{equation}
	\mathbb{E}_{p_{m}^{(j-1)}}[\mbu] = \mbm^{(j-1)} + \gamma\Delta\mbm^{(j - 2)},
\end{equation}
where $\gamma (< 1)$ is a decay parameter.
By replacing $p^{(j-1)}$ in Eq.~(\ref{eqn:update_law_p}) to $p_{m}^{(j-1)}$, we get an update law considering the momentum:
\begin{equation}
  \begin{split}
  	\mbm^{(j)} &= \mbm^{(j-1)} + \Delta\mbm^{(j-1)}, \\
  	\Delta\mbm^{(j-1)} &= \gamma \Delta\mbm^{(j-2)} + \delta\mbm^{(j-1)}, \\
  	\delta\mbm^{(j-1)} &= \frac{\mathbb{E}_{p_{m}^{(j-1)}}\left[e^{- S(\tau) / \lambda}\mbe\right]}{\mathbb{E}_{p_{m}^{(j-1)}}\left[e^{- S(\tau)/ \lambda}\right]}. \label{eqn:update_law_nag}
  \end{split}
\end{equation}
After a few modifications of Alg.~\ref{alg:pi}, the implementation of (\ref{eqn:update_law_nag}) is achieved, as shown in Alg.~\ref{alg:pi_nag} where modified or added lines are highlighted.
Fig.~\ref{fig:nag} shows an intuitive illustration of this approach.

\subsubsection{AdaGrad \cite{duchi2011adaptive}}
AdaGrad adapts step-sizes considering the accumulation of past gradients.
To apply this concept to our case, we modify the top equation of (\ref{eqn:update_law_nag}):
\begin{equation}
	\mbm^{(j)} = \mbm^{(j-1)} + {\boldsymbol \eta}^{(j-1)} \circ \Delta\mbm^{(j-1)},
\end{equation}
where ${\boldsymbol \eta}^{(j-1)} \in \mathbb{R}^{m \times T}$ is the newly introduced step size vector, and ``$\circ$'' indicates the element-wise product.
Starting with ${\boldsymbol \eta}^{(-1)} = \mathbf{1}$, ${\boldsymbol \eta}^{(j-1)}$ is adapted by the update law of AdaGrad. 
% Learning to learn \cite{tamar2016learning}.

\subsubsection{Adam \cite{kingma2014adam}}
Adam is the combination of the momentum-based acceleration and step-size adaptation.
Our implementation uses the same equations and hyper-parameters proposed in the reference.

The above introductions are rather heuristic and we will not discuss their effect on the convergence from a theoretical perspective.
Contrary, another accelerated mirror descent method that assures the rate of $O(1/j^{2})$ is theoretically derived~\cite{krichene2015accelerated, miyashita2017mirror}.
However, in this study, we did not employ said method because some equations in it cannot be represented as a closed-form.
Therefore, additional iterative algorithms \cite{krichene2015efficient} must be introduced into the iterative path integral method to solve the equations, making it difficult to apply this method to real-time MPC and PI-Net.

%%%% Pseudo Code of SOC %%%%
\begin{algorithm}[tbh]
\small
\caption{\small NAG-Accelerated Path Integral Method} \label{alg:pi_nag}
\begin{algorithmic}[1]
% \INPUT $K$, $T$: Number of trajectories \& timesteps \\
\Require $\mbx_{0}$, $\mbm^{(j-1)}$: Input state \& Control sequence
\Statex \underline{$\Delta\mbm^{(j-2)}$: Momentum} 
\Ensure $\mbm^{(j)}$: Improved control sequence 
\Statex \underline{$\Delta\mbm^{(j-1)}$: Momentum}

% Initialization
\State Sample $\{\mbe^{(0)}, \mbe^{(1)}, \cdots, \mbe^{(K-1)}\}$
% \State $\mathcal{T} \leftarrow \emptyset$

% Monte-Carlo simulation
\For{$k \leftarrow 0$ \textbf{to} $K-1$}
\State $\mbx^{(k)}_{0} \leftarrow \mbx_{0}$
\State $\tau^{(k)} \leftarrow \emptyset$
\For{$t \leftarrow 0$ \textbf{to} $T-1$}
\State $\tau^{(k)} \leftarrow \tau^{(k)} \cap \{\mbx^{(k)}_{t+1}, \mbu^{(k)}_{t}\}$
\State \underline{$\mbu^{(k)}_{t} \leftarrow \mbm^{(j-1)}_{t} + \gamma \Delta\mbm_{t}^{(j-2)} + \mbe^{(k)}_{t}$}
\State $\mbx^{(k)}_{t + 1} \leftarrow f(\mbx^{(k)}_{t}, \mbu^{(k)}_{t})$
\EndFor
\State $\tau^{(k)} \leftarrow \tau^{(k)} \cap \{\mbx^{(k)}_{T}\}$
% \State $\mathcal{T} \leftarrow \mathcal{T} \cap \{\tau^{(k)}\}$
\EndFor

% Update control sequence
\State \underline{$\delta\mbm^{(j-1)} \leftarrow {\sum^{K-1}_{k=0}e^{-S(\tau^{(k)})/\lambda}\mbe^{(k)}} \big/ {\sum^{K-1}_{k=0}e^{-S(\tau^{(k)})/\lambda}}$}
\State \underline{$\Delta\mbm^{(j-1)} \leftarrow \gamma \Delta\mbm^{(j-2)} + \delta\mbm^{(j-1)}$}
\State $\mbm^{(j)} \leftarrow \mbm^{(j-1)} + \Delta\mbm^{(j-1)}$
\end{algorithmic}
\end{algorithm}

%%%%%%%%%%%%%%%%%%%%%%%%%%%%

\begin{figure*}[t]
    \centering
    \includegraphics[width=0.8\textwidth]{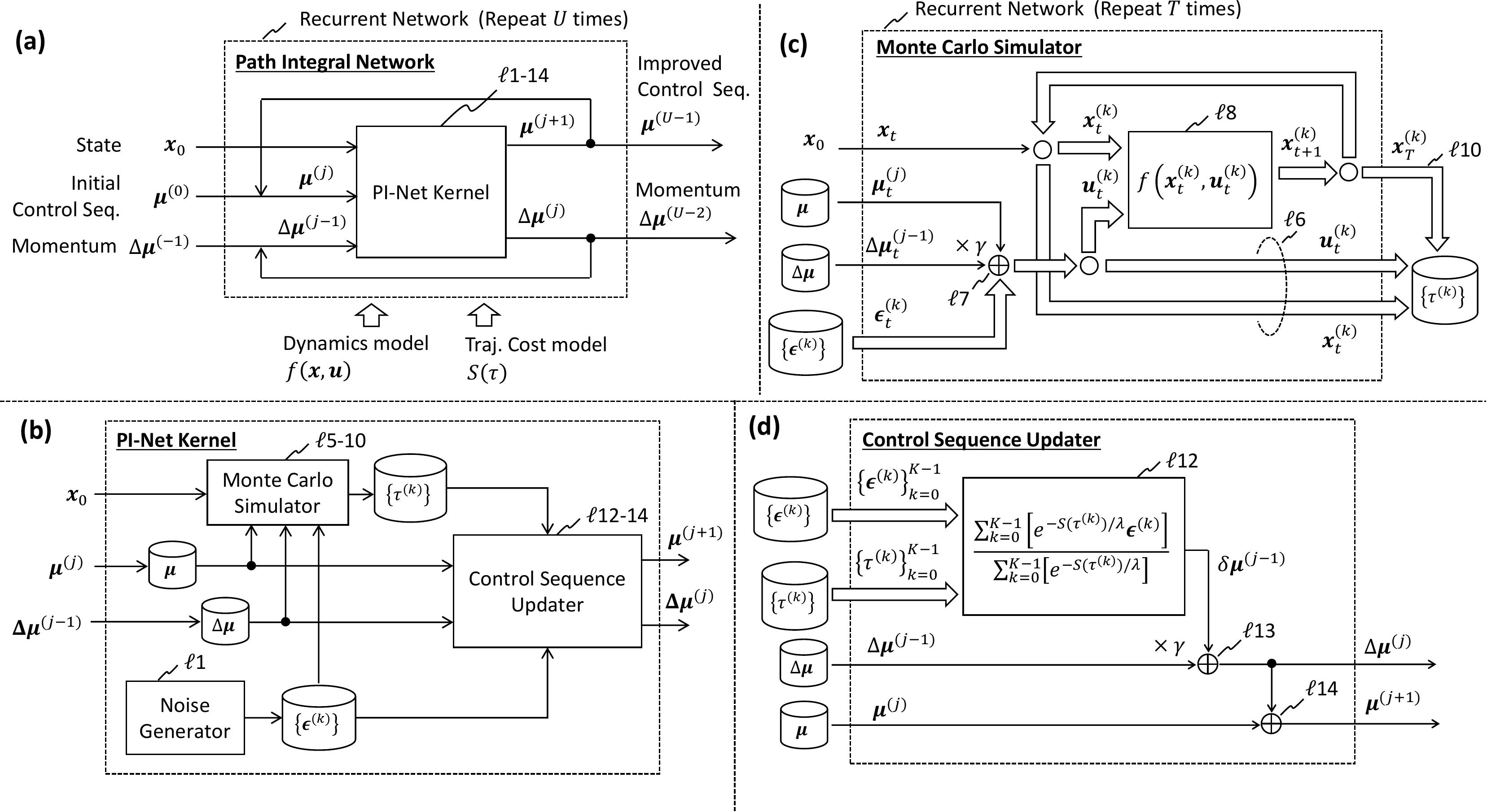}
    \caption{
    	Architecture of the accelerated PI-Net. `$\ell$' labels indicate the corresponding line numbers in Alg.~\ref{alg:pi_nag}.
    	Block arrows in (c,d) indicate multiple signal flow with respect to $K$ trajectories.} \label{fig:pi_arch}
\end{figure*}

\subsection{Application to Model Predictive Control}
This section and the next focus on the NAG accelerated path integral method and exemplify its applications to MPC and PI-Net.

Alg.~\ref{alg:mpc} shows the procedure of MPC using the accelerated path integral.
In $\ell$7--8, not only the control sequence $\mbm_{1:T-1}$,
but also the momentum $\Delta\mbm_{1:T-1}$ are retained for the next step to \textit{warm start} the optimization.
% The effect of this scheme is demonstrated in the experiment in Sect.~\ref{ssec:exp_mpc}.
% Retaining of $\Delta\mbm_{1:T-1}$ is not always conducted; we reset $\Delta\mbm_{1:T-1}$ as zero-vector in $\ell 10$ when a reset condition is satisfied (e.g.~sudden situation change).
In our implementation, $\ell 9$ is performed as $\mbm_{T-1} \leftarrow \mbm_{T-2}$ and $\Delta\mbm_{T-1} \leftarrow \Delta\mbm_{T-2}$.
%
%%%% Pseudo Code of MPC %%%%
\begin{algorithm}[tbh]
\small
\caption{\small MPC with NAG-Accelerated Path Integral} \label{alg:mpc}
\begin{algorithmic}[1]
\Require $\mbm$, $\Delta\mbm$: Initial control sequence \& Momentum 
% Initialization
\While {True}
	\State Observe current state $\mbx_{t}$
    \For{$j \leftarrow 0$ \textbf{to} $U$}
    	\State $(\mbm, \Delta\mbm) \leftarrow$ Execute Alg.~\ref{alg:pi_nag} inputting $(\mbx_{t}, \mbm, \Delta\mbm)$
    \EndFor
    \State Send $\mbm_{0}$ to actuators
    \State $\mbm_{0:T-2} \leftarrow \mbm_{1:T-1}$
	\State $\Delta\mbm_{0:T-2} \leftarrow \Delta\mbm_{1:T-1}$
	\State Initialize $\mbm_{T-1}$, $\Delta\mbm_{T-1}$
%     \State Initialize 
% 	\State Initialize $\mbm_{T-1}$
% %     \State $\mbm_{T-1} \leftarrow \mbm_{T-2}$
% %     \State $\Delta\mbm_{T-1} \leftarrow \Delta\mbm_{T-2}$
%     \IF{Reset condition for $\Delta\mbm_{T-1}$ is satisfied}
%     	\State $\Delta\mbm_{0:T-2} \leftarrow \mathbf{0}$
%     \ELSE
%     	\State $\Delta\mbm_{0:T-2} \leftarrow \Delta\mbm_{1:T-1}$
%         \State Initialize $\Delta\mbm_{T-1}$
%     \ENDIF
\EndWhile
\end{algorithmic}
\end{algorithm}

%%%%%%%%%%%%%%%%%%%%%%%%%%%%

%  1752  python make_training_data.py -R1 10 -R2 10 -K 100 --pi --gamma 0.8 --npz icra_g1_s --thd_d 0.05 --thd_q 0.1 --thd_s 1. -j 5 --tolFun -1
%  1753  python make_training_data.py -R1 10 -R2 10 -K 100 --pi --gamma 0.0 --npz icra_g0_s --thd_d 0.05 --thd_q 0.1 --thd_s 1. -j 5 --tolFun -1
%  1758  python make_training_data.py -R1 10 -R2 10 -K 100 --pi --gamma 0.8 --npz icra_g1_L10_s --length 10
%  1759  python make_training_data.py -R1 10 -R2 10 -K 100 --pi --gamma 0. --npz icra_g0_L10_s --length 10

\subsection{Application to PI-Net and Inverse Optimal Control}
Fig.~\ref{fig:pi_arch} illustrates the network representation of Alg.~\ref{alg:pi_nag}.
The operations of this network are essentially the same than in the original PI-Net described in reference \cite{okada2017path}, except for the flows of the momentum $\Delta\mbm$, which are newly added.
Thus we omit a detailed explanation of how this network operates.
This network is certainly differentiable and trained by standard back-propagation.
We can hence utilize the same training schemes than those of the original PI-Net to carry out the inverse optimal control.

\section{Experiments} \label{sec:exp}
This section presents several experiments that we conducted under simulated settings to examine the performance of the accelerated path integral method on convergence, MPC, and inverse optimal control.

\subsection{Convergence Rate Comparison} \label{ssec:exp_convergence}
The objective of this experiment was to evaluate the convergence performance of the iterative path integral combined with the gradient methods from Sect.~\ref{sec:accel}.
We also compared the path integral method with following methods:
1) Differential Dynamic Programming (DDP) \cite{todorov2005generalized} as a reference to demonstrate the validity of our method and its implementation, and
2) the original iterative path integral method \cite{williams2017information} as a baseline to examine how adapting gradient methods could improve the performance.

\begin{figure*}[t]
	\centering
    \subfloat[Inverted Pendulum \cite{wang1996approach}]{
    \hspace{1cm}
    \includegraphics[height=3.5cm]{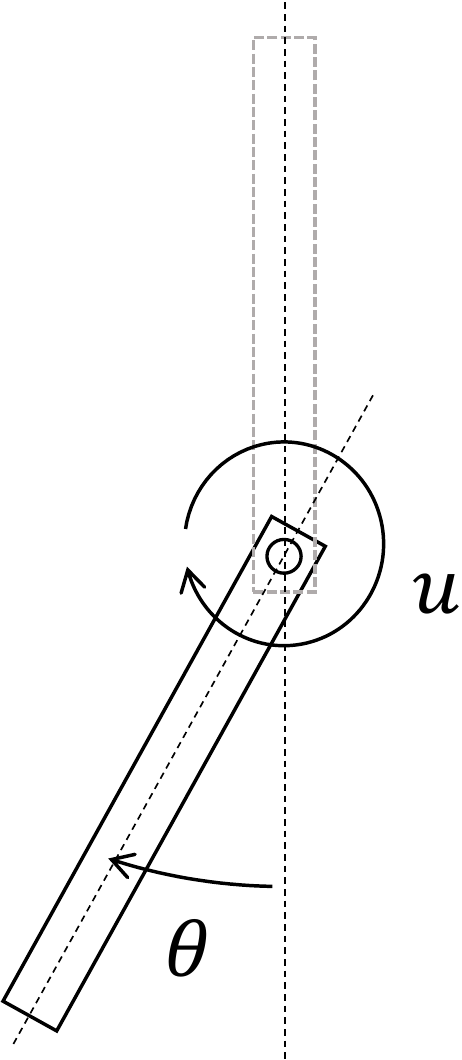}
    \hspace{1cm}
    }
    \subfloat[Hovercraft \cite{seguchi2003nonlinear}]{
    	\includegraphics[height=3.5cm]{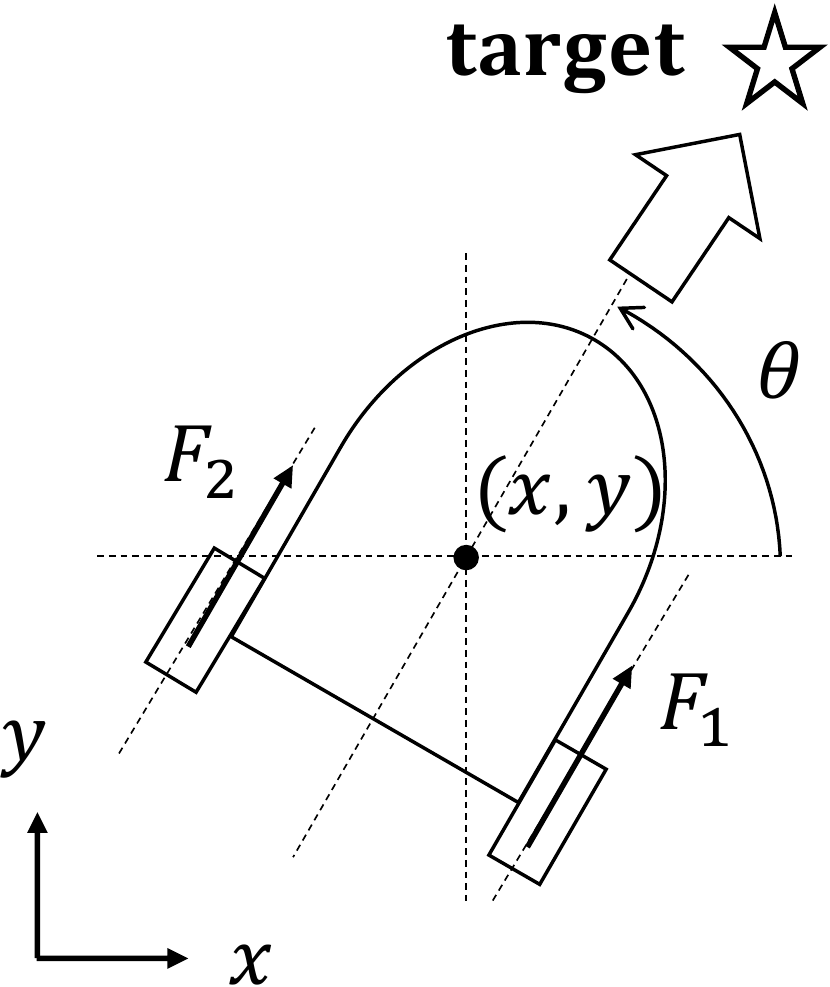}}
    \quad\quad
    \subfloat[Quadrotor \cite{michael2010grasp}]{
    	\includegraphics[height=2.8cm]{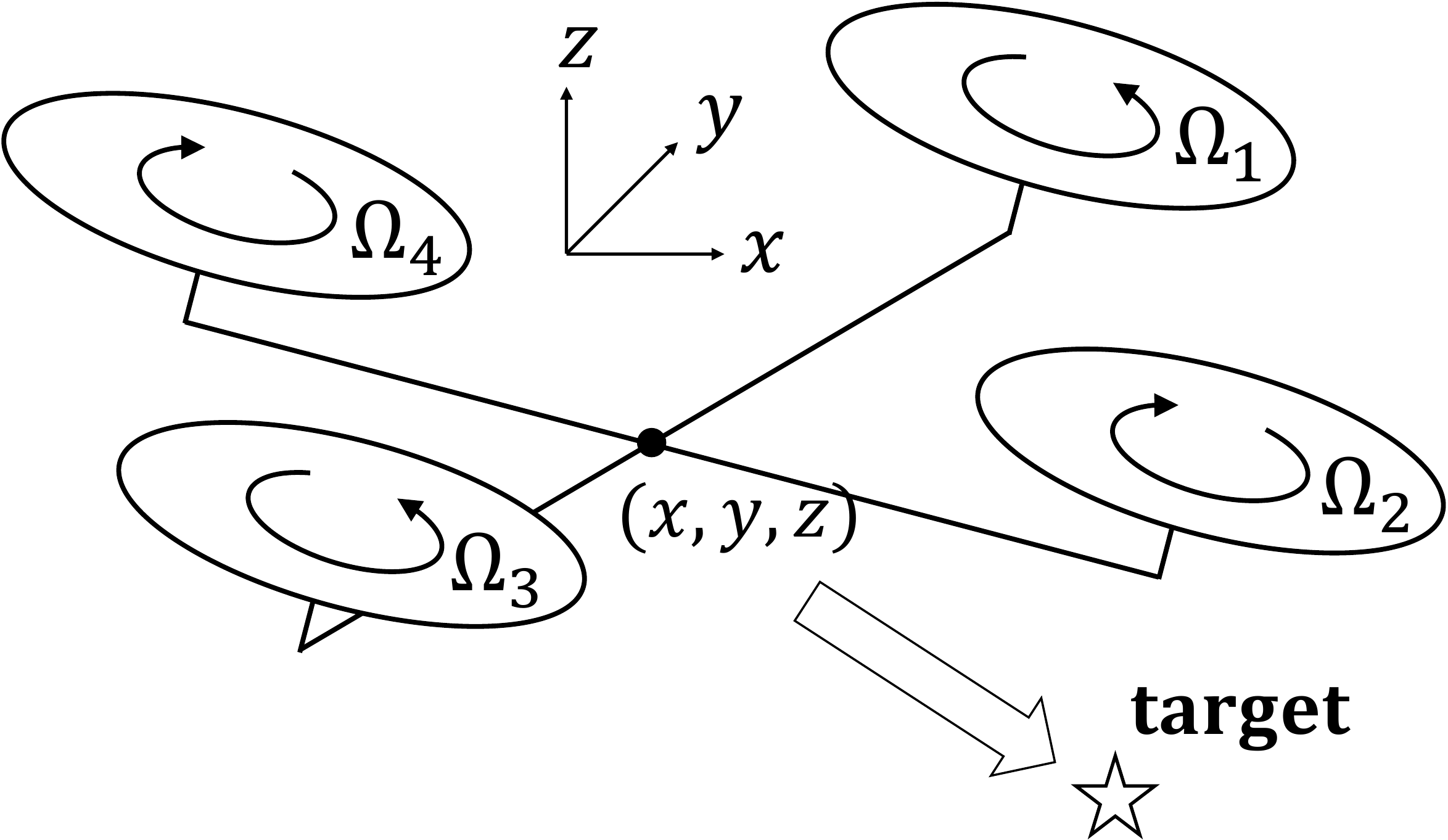}}
    \quad\quad
    \subfloat[Car (bicycle model) \cite{gonzales2016autonomous}]{
    	\includegraphics[height=3.5cm]{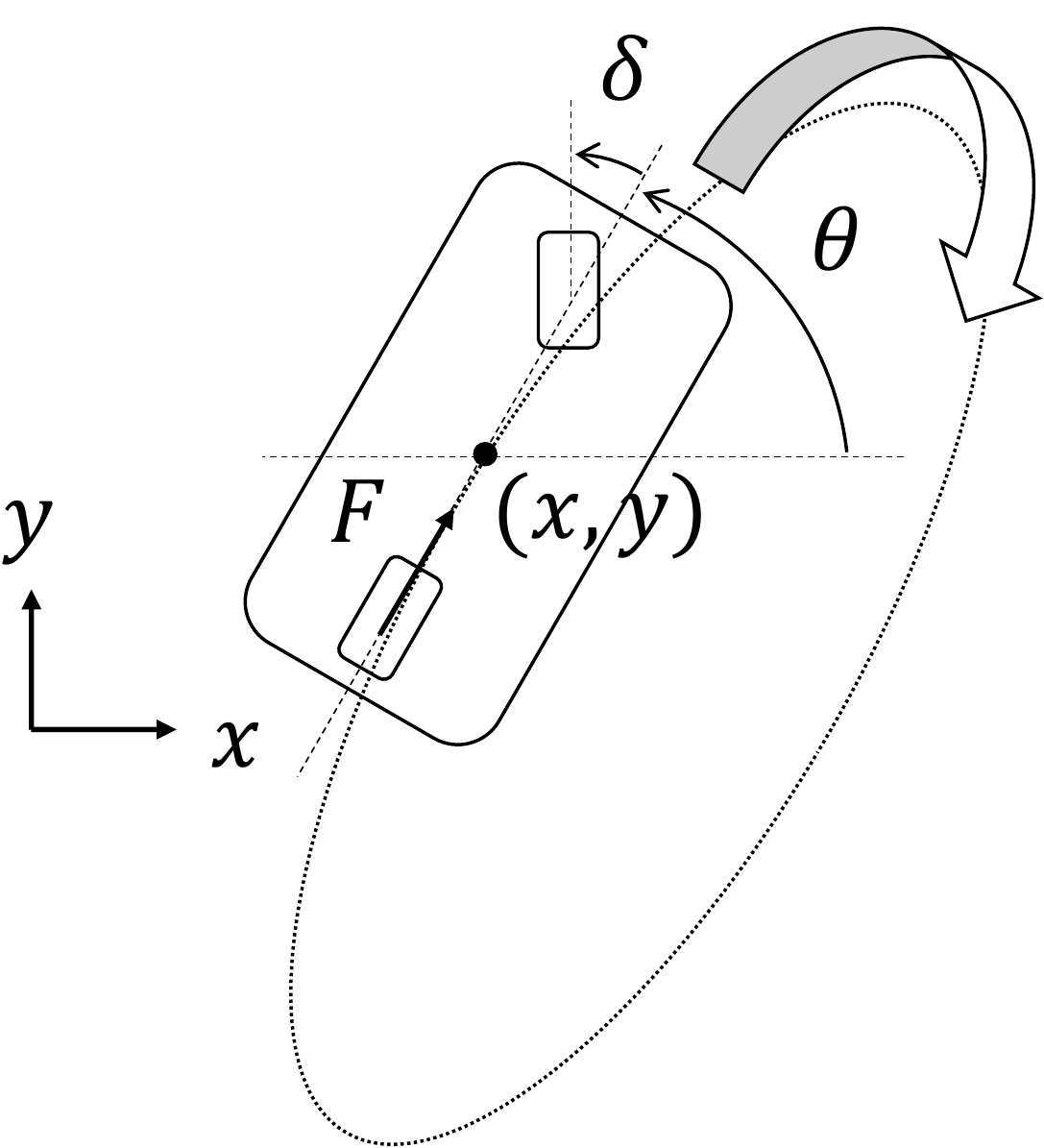}}
    \caption{
    Sketches of the systems and their control tasks. 
    States of each system comprises positions, orientations and their time-derivatives. The following are also regarded as states: current thruster outputs $F_{1,2}$ (hovercraft), current rotor frequencies ${\boldsymbol \Omega} = \Omega_{1,2,3,4}$ (quadrotor), and current steering angle $\delta$ and rear-tire force $F_{r}$ (car). Control inputs are torque $u$ (inverted-pendulum) and the desired values of $F_{1,2}$,  ${\boldsymbol \Omega}$, $(\delta, F_{r})$.
    } \label{fig:vehicle_systems}
\end{figure*}

We focused on four non linear simulated dynamics systems: inverted pendulum \cite{wang1996approach}, and miniature vehicle systems (hovercraft \cite{seguchi2003nonlinear}, quadrotor \cite{michael2010grasp} and car \cite{gonzales2016autonomous}).
The aims of the control tasks are as follows:
to swing the pendulum up (inverted pendulum), to navigate the vehicles towards target positions (hovercraft and quadrotor) and to navigate the vehicle to go around an oval track at a desired speed.
Fig.~\ref{fig:vehicle_systems} illustrates the systems and their control tasks.
The cost functions that encode the above tasks are summarized in Appx.~\ref{app:cost}.
Parameters of the path integral method are commonly set as $\lambda = 0.01$ and $K=1000$.
Initial control sequence and momentum were initialized as $\mbm^{(0)} = \mathbf{0}$ and $\Delta\mbm^{(-1)} = \mathbf{0}$.
The decay parameter of NAG was set as $\gamma = 0.8$. 

\begin{figure*}[t]
    \centering
    \includegraphics[width=0.245\textwidth]{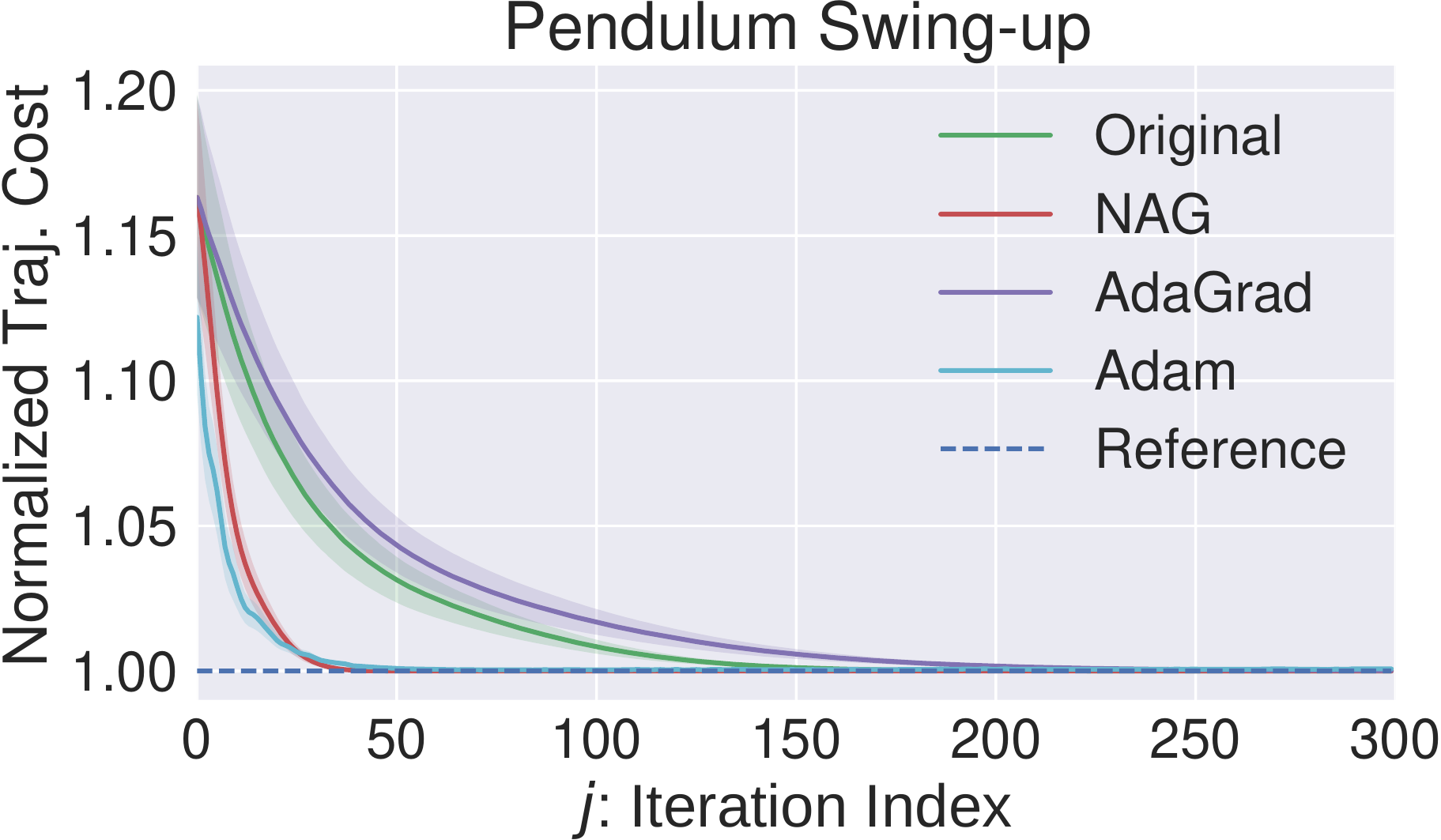}
    \includegraphics[width=0.245\textwidth]{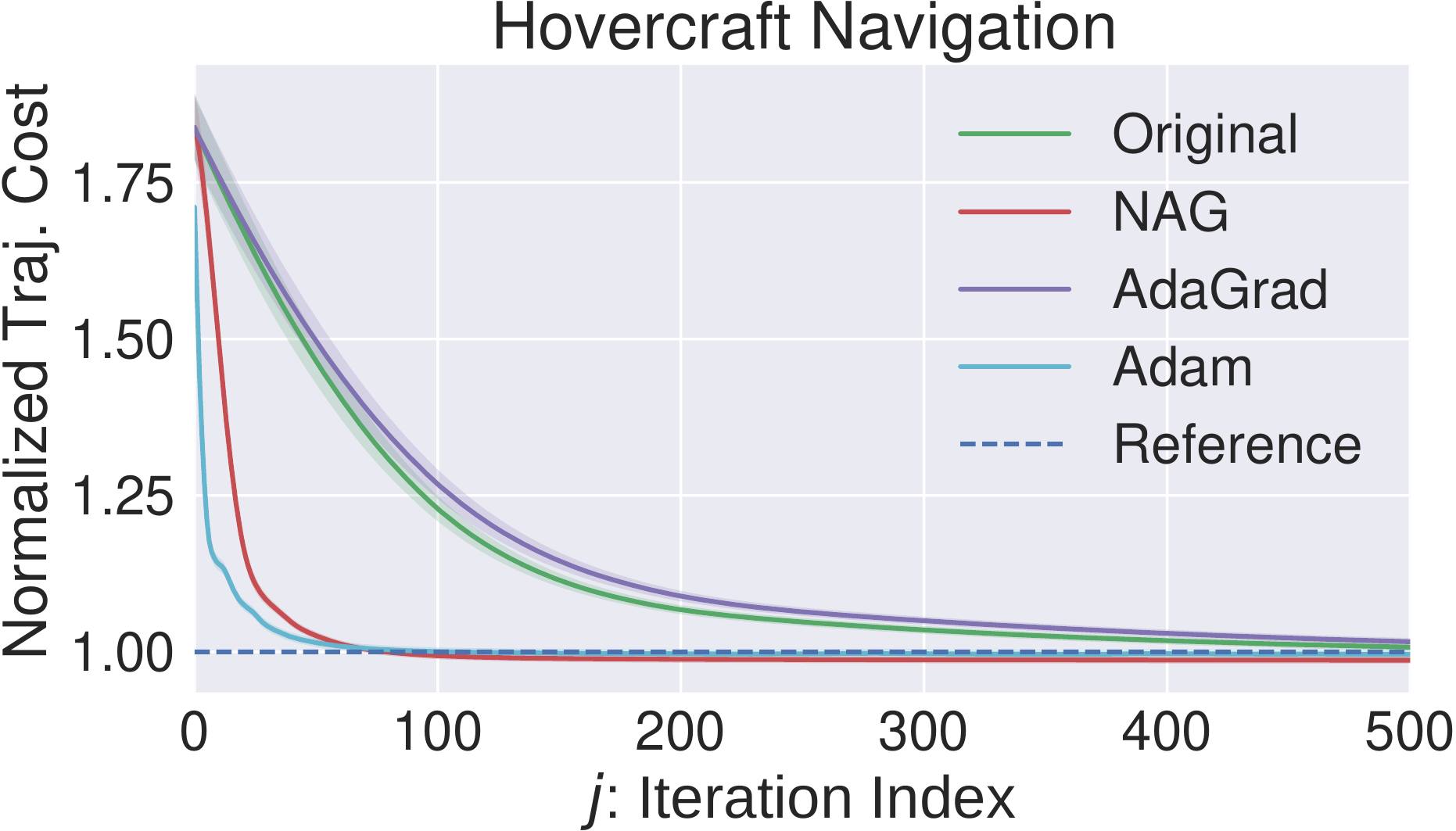} 
    \includegraphics[width=0.245\textwidth]{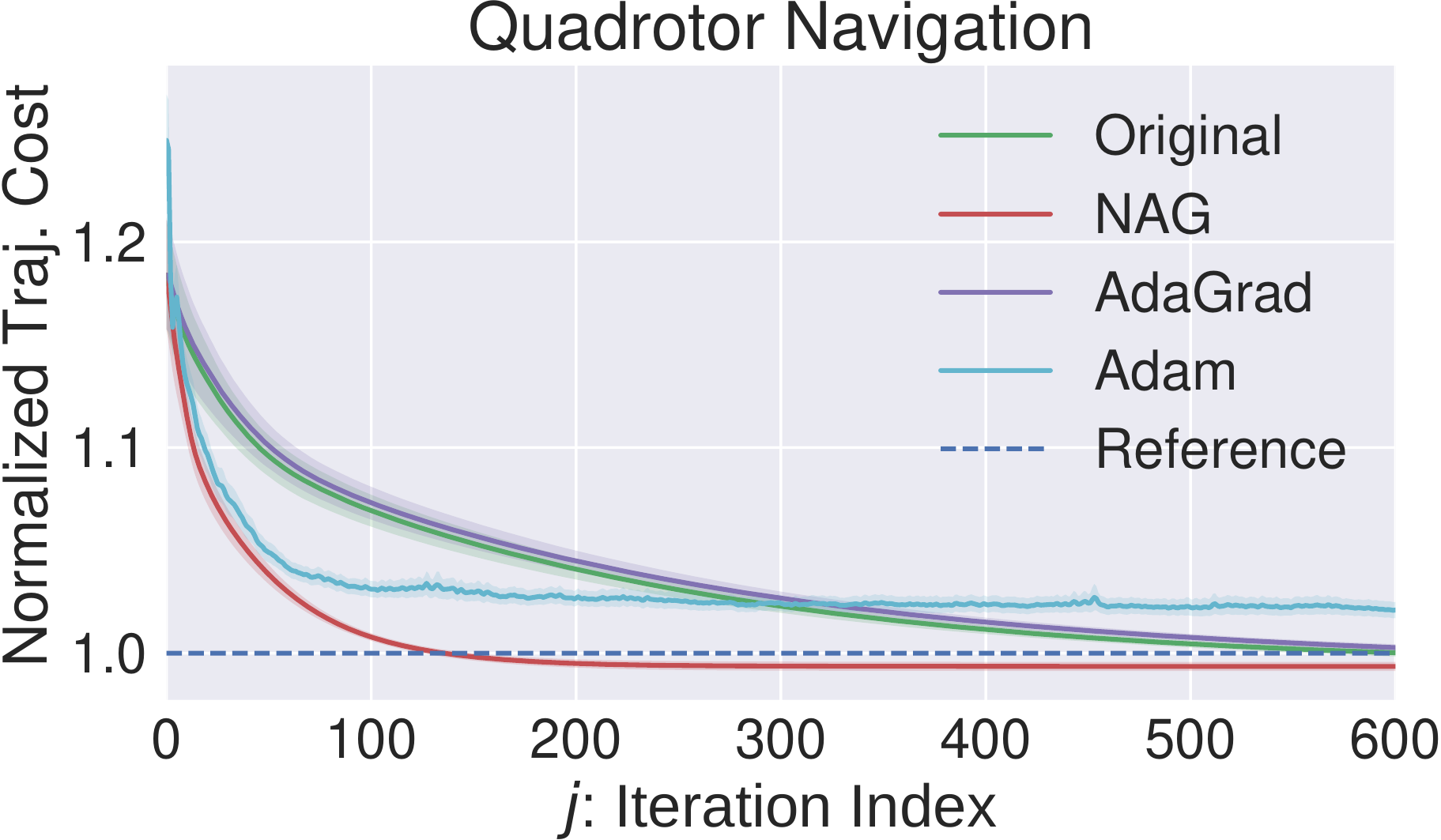}
    \includegraphics[width=0.245\textwidth]{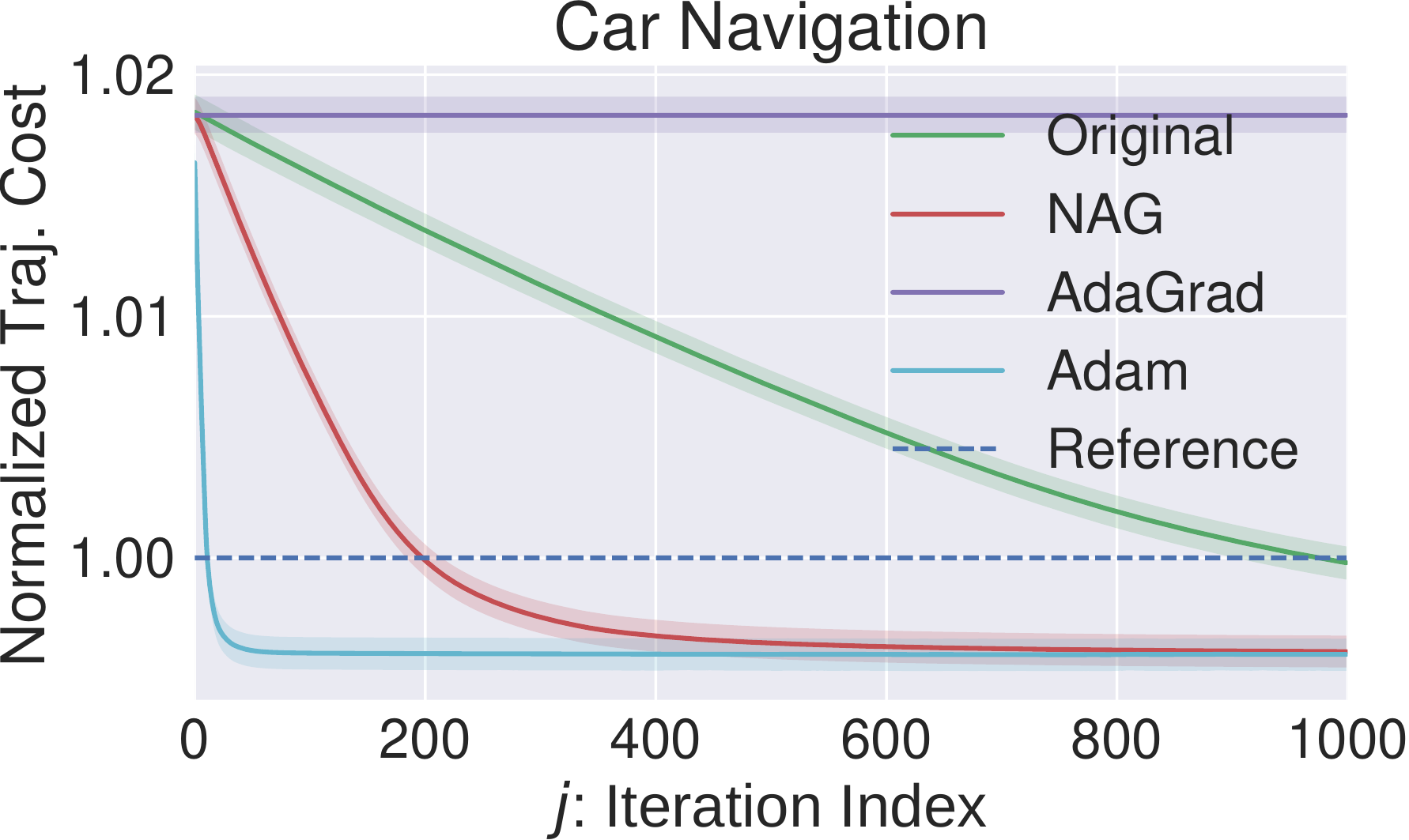}
    \caption{Results of the experiment on convergence performance from Sect.~\ref{ssec:exp_convergence}. For each setting, we conducted 100 optimizations with different state inputs and the results represent the average. In order to verify the optimization results, we also show reference cost values obtained from DDP. The cost values were normalized ($\mathrm{references} = 1$)} \label{fig:convergence}
\end{figure*}

With these conditions, we observed the convergence performance via optimization of the cost functions.
Fig.~\ref{fig:convergence} summarizes the results.
In this figure, NAG-accelerated path integral achieved faster convergence than the baseline for all control tasks. 
In addition, the converged cost values are lower than the reference cost by DDP for the three navigation tasks.
Although Adam resulted in unstable convergence in the quadrotor task, it achieved much faster convergence than NAG for other tasks particularly for the car navigation task.
In this experiment, AdaGrad showed no improvement from the baseline.

\subsection{Model Predictive Control} \label{ssec:exp_mpc}
\begin{figure*}[t]
    \centering
    \includegraphics[width=0.3\textwidth]{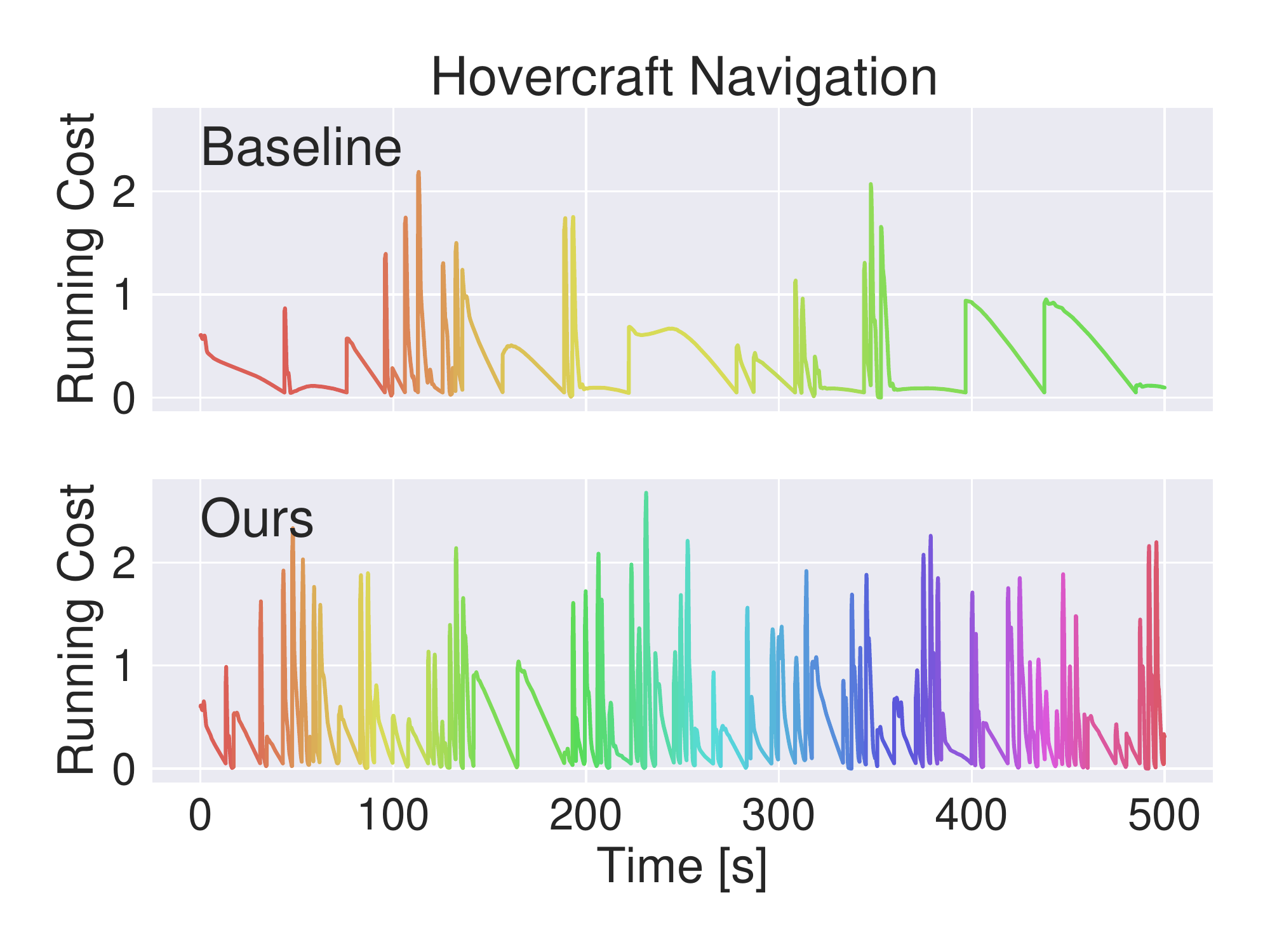}
    \includegraphics[width=0.3\textwidth]{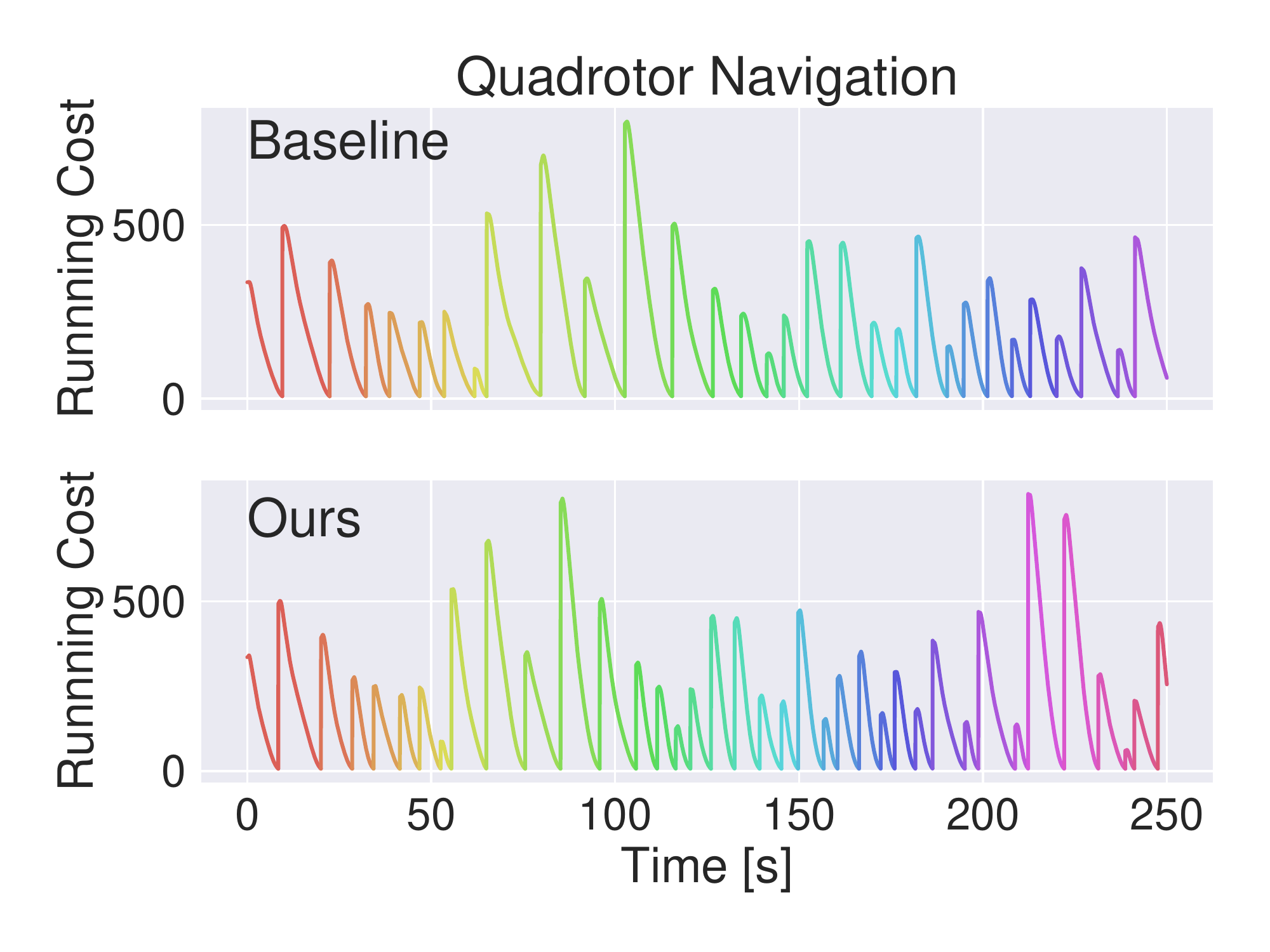}
    \includegraphics[width=0.3\textwidth]{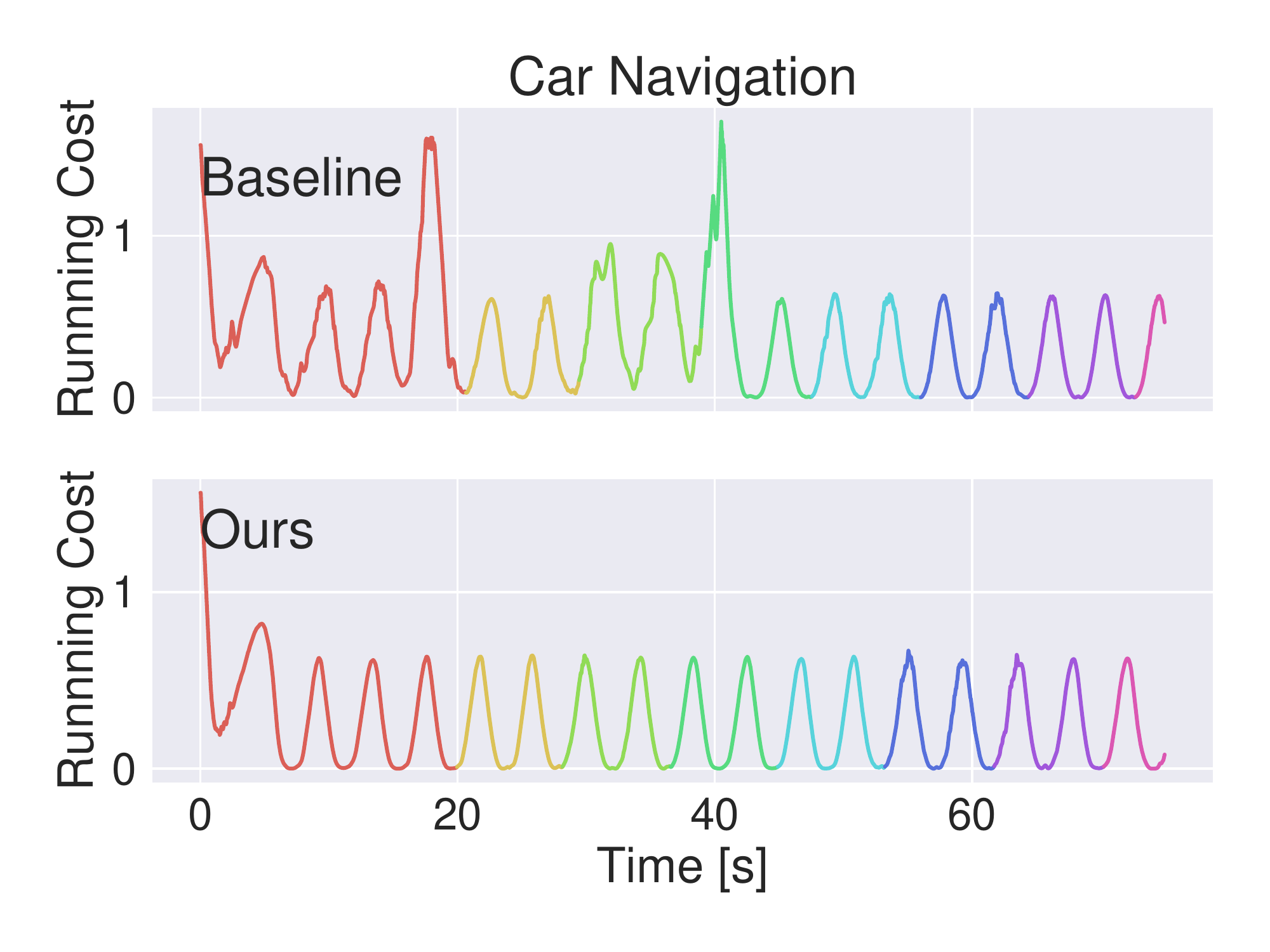}
    \vspace{-5mm}
    \caption{
    Time transition of running cost $q(\mbx)$ during the simulations of vehicle navigations.
	For clarity purposes, these figures focus only on the first seconds of the entire simulations.
    Changes of color indicate task completions; i.e., faster color change means faster task completions.
    } \label{fig:mpc_result}
\end{figure*}

We conducted MPC simulations of the three vehicle systems with Alg.~\ref{alg:mpc} and a baseline method (i.e., MPC with the original path integral \cite{williams2017information}).
The cost functions mentioned above were also used in this section.
In the simulations of the hovercraft and quadrotor, a target position was changed when tasks were completed (i.e., vehicle reached to the target).
Path integral parameters were set as $(U, K, \gamma) = (10, 100, 0.8)$ for the hovercraft and quadrotor navigation tasks and $(U, K,\gamma) = (25, 200, 0.3)$ for the car navigation task.
The results of these simulations are summarized in Fig.~\ref{fig:mpc_result} and Table~\ref{tab:mpc_result}
in which our accelerated method successfully improved control performances especially for the hovercraft task.
% In the car navigation task, both of the controller soon 
% The improvement on the car navigation task, which is rather periodical control task, was less significant compared to the others.
%
\begin{table*}[t]
  \centering
  \caption{Summary of MPC simulation results.} \label{tab:mpc_result}
%   \begin{tabular}{c|cc|cc}
%   	\toprule
% 	& \multicolumn{2}{c|}{Hovercraft} & \multicolumn{2}{c}{Quadrotor} \\\hline
%     & Baseline \cite{williams2017information} & Ours & Baseline  \cite{williams2017information} & Ours \\\hline
%     Number of Task Completions & $68$ & $\mathbf{220}$ & $87$ & $\mathbf{108}$ \\
%     Mean Time to Task Completion [s] & $17.93$ & $\mathbf{5.700}$ & $8.713$ & $\mathbf{7.004}$ \\
%     Mean Accumulated-cost to Task Completion & $237.5$ & $\mathbf{91.92}$ & $5.386 \times 10^{4}$ & $\mathbf{3.849 \times 10^{4}}$ \\
  \begin{tabular}{c|cc|cc|cc}
  	\toprule
	& \multicolumn{2}{c|}{Hovercraft} & \multicolumn{2}{c|}{Quadrotor} & \multicolumn{2}{c}{Car} \\\hline
    & Baseline \cite{williams2017information} & Ours & Baseline  \cite{williams2017information} & Ours & Baseline  \cite{williams2017information} & Ours \\\hline
    Number of Task Completions (\# Reached-Targets or \# Laps) & $68$ & $\mathbf{220}$ & $87$ & $\mathbf{108}$ & $\mathbf{8}$ & $\mathbf{8}$ \\
    Mean Time to Task Completion [s] & $17.93$ & $\mathbf{5.700}$ & $8.713$ & $\mathbf{7.004}$ & $9.094$ & $\mathbf{8.784}$ \\
    Mean Accumulated-cost to Task Completion & $237.5$ & $\mathbf{91.92}$ & $5.386 \times 10^{4}$ & $\mathbf{3.849 \times 10^{4}}$ & $132.8$ & $\mathbf{91.86}$ \\
  	\bottomrule
  \end{tabular}
\end{table*}

\subsection{Inverse Optimal Control}
Inverse optimal control was conducted with both the original PI-Net \cite{okada2017path} and the accelerated one.
We focused on the pendulum swing-up task described above.
We considered DDP with the given dynamics and cost models as an expert and generated demonstrations in the same manner as in reference \cite{okada2017path}.
From the demonstrations, we prepared two dataset $\mcsymbol{D}{train}$ and $\mcsymbol{D}{test}$, each of which was used for training or test.
The dataset takes the form $(\mbx^{\star}_{t}, \mbm^{\star}_{t}) \in \mathcal{D}$ where $\mbx^{\star}_{t}$ is a state input and $\mbu^{\star}_{t}$ is the corresponding control by the expert.

The dynamics in PI-Net was same as with the expert and we represent the cost model as a neural network, which had a single hidden layer with 12 hidden nodes and arctangent activation functions.
Then, PI-Net was trained by optimizing the mean squared errors (MSE) between the PI-Net output $\mbm^{(U-1)}_{0}$ and the expert control $\mbm^{\star}_{t}$ so that the neural cost approximates the true cost.
Common parameters were set as $K=100$, $U \in \{50, 100, 200\}$.
The decay parameter $\gamma$ was set to $0.8$.

\begin{figure}[t]
    \centering
    \includegraphics[width=0.5\textwidth]{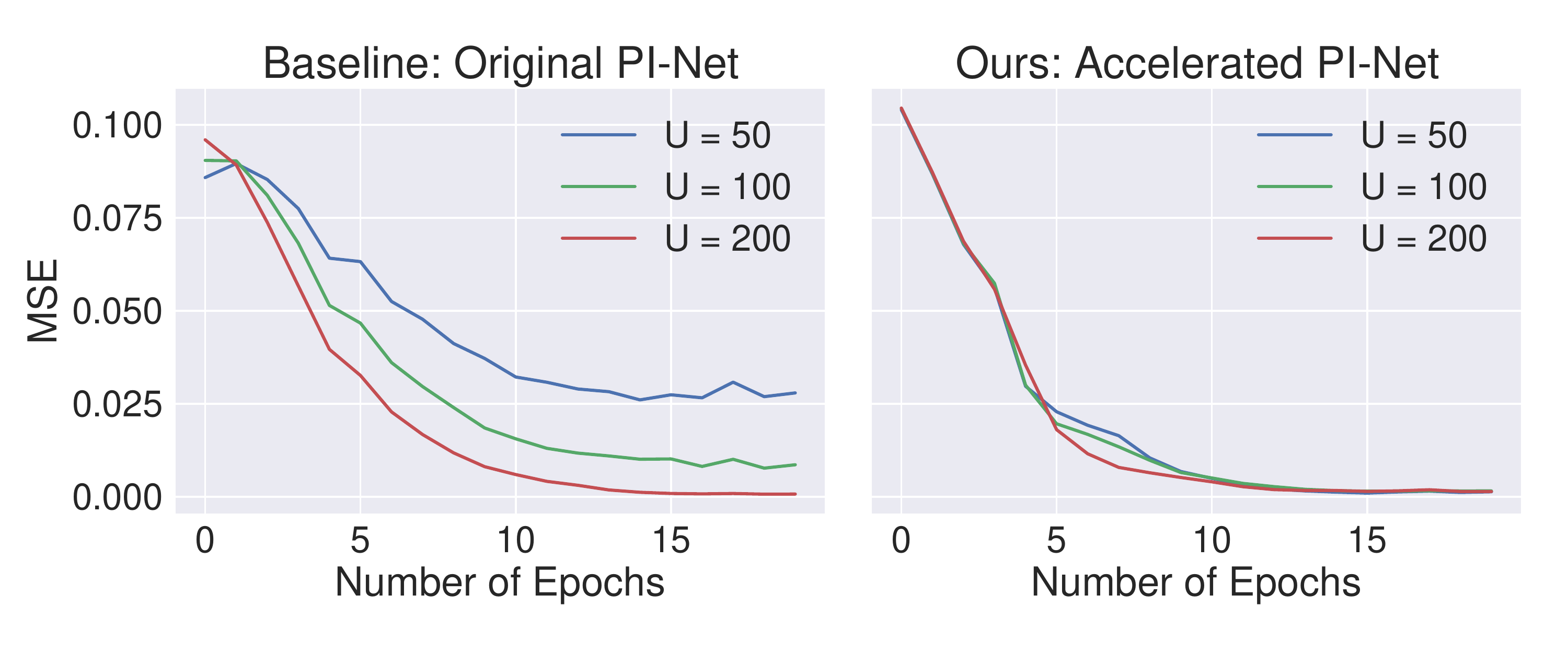}
    \vspace{-7.5mm}
    \caption{Convergence of MSE w.r.t. $\mcsymbol{D}{train}$ during PI-Net traning.} \label{fig:exp_pi_net}
\end{figure}

Fig.~\ref{fig:exp_pi_net} shows the convergence of the MSEs during training,
where the small number of iterations (i.e.~$U=\{50, 100\}$) impeded sufficient convergence for the baseline.
In contrast, accelerated PI-Net results show efficient convergences for all $U$s.
This result contributes to significantly reduce the computational complexity for back-propagation as summarized in Table~\ref{tab:ram_usage}.
\begin{table}
	\centering
    \caption{RAM Usage and Computational Time for PI-Net training} \label{tab:ram_usage}
    \begin{tabular}{c|ccc}
    	\toprule
        $U$ & 50 & 100 & 200 \\\hline
        RAM [GB] & 32.5 & 49.9 & 81.3 \\
        Time per Epoch [s] & 335 & 653 & 1318 \\
        \bottomrule
    \end{tabular}
\end{table}

\begin{figure}[t]
    \centering
    \includegraphics[width=0.5\textwidth]{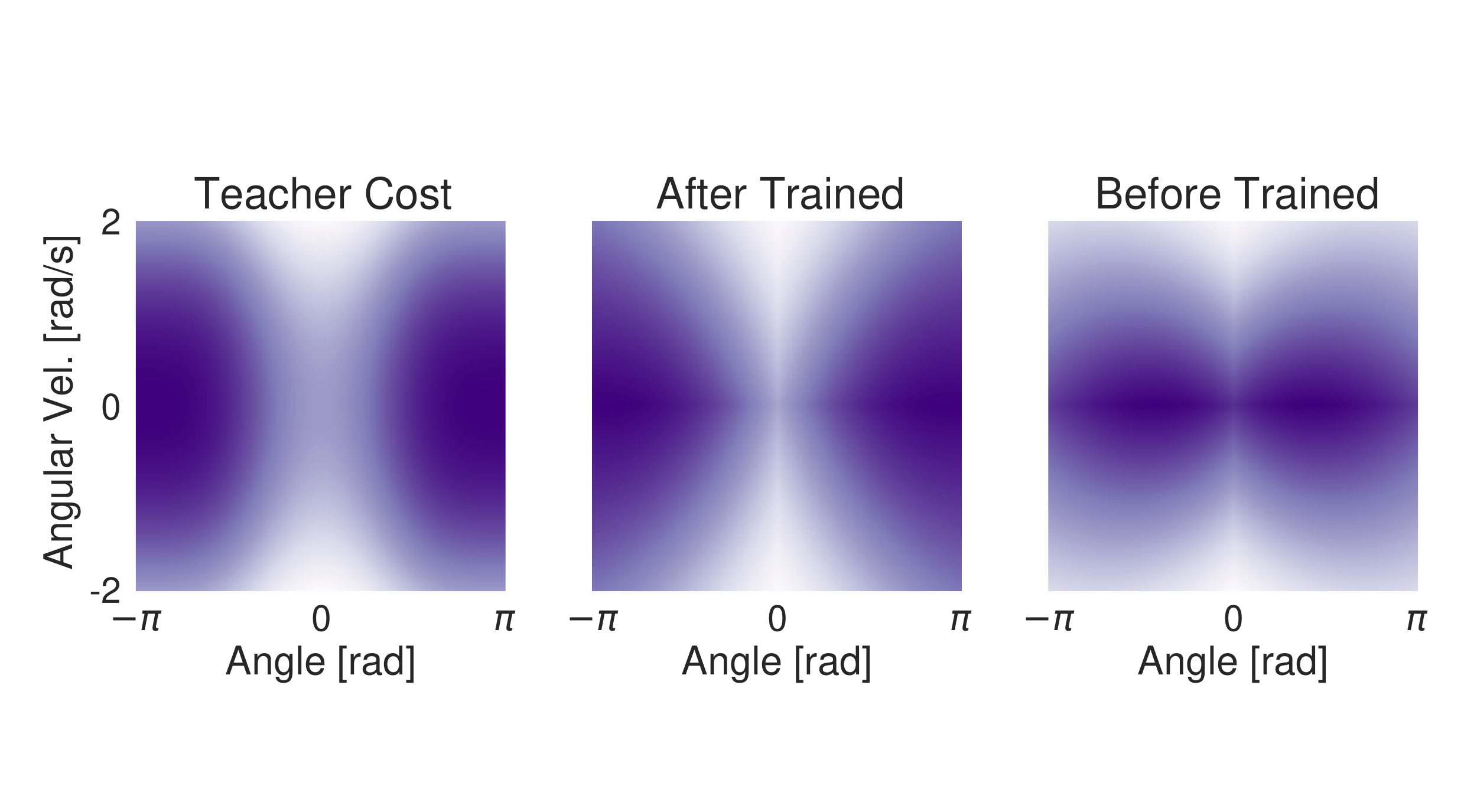}
    \vspace{-10mm}
    \caption{Visualized cost map from PI-Net training.} \label{fig:cost_map}
\end{figure}

Table.~\ref{tab:exp_pi_net} summarizes the training and test errors between the reference and the output of trained PI-Net with different $U$s.
The accelerated PI-Net leads to smaller errors when $U \in \{50, 100\}$, however,
the baseline with no acceleration shows better results for $U = 200$.
We suppose this results from the side effect of the momentum-based acceleration; the accumulated past momenta interfered with \textit{fine} optimization.
We believe that carefully scheduling the decay parameter $\gamma$ over $U$ iterations can alleviate this.
Fig.~\ref{fig:cost_map} illustrates the learned cost model by the accelerated PI-Net ($U=200$).
\begin{table}[t]
  \centering
  \caption{MSE b/w Trained PI-Net Outputs and the Expert Demonstrations.}\label{tab:exp_pi_net}
  \begin{tabular}{c|c|cc}
  	\toprule
	    & $U$ & Baseline \cite{okada2017path} & Ours \\\hline
                            & 50 & $2.736 \times 10^{-2}$ & $\mathbf{1.234 \times 10^{-3}}$ \\
       $\mcsymbol{D}{train}$ & 100 & $8.775 \times 10^{-3}$ & $\mathbf{1.670 \times 10^{-3}}$ \\
                            & 200 & $\mathbf{6.625 \times 10^{-4}}$ & $1.720 \times 10^{-3}$ \\\hline
                            & 50 & $1.788 \times 10^{-2}$ & $\mathbf{1.613 \times 10^{-3}}$ \\
       $\mcsymbol{D}{test}$ & 100 & $4.523 \times 10^{-3}$ & $\mathbf{1.302 \times 10^{-3}}$ \\
                            & 200 & $\mathbf{4.630 \times 10^{-4}}$ & $1.340 \times 10^{-3}$ \\
  	\bottomrule
  \end{tabular}
\end{table}

\section{Conclusions}
In this work, we developed a gradient-based and accelerated iterative path integral optimal control method.
This work is greatly inspired by the mirror descent search from Miyashita et al.~\cite{miyashita2017mirror}, which is such a powerful and general framework that one can directly apply it to the stochastic optimal control problem, thus deriving a gradient-based iterative path integral method.
Given the relation between the path integral and gradient descent, we could introduce optimization methods used for gradient descent into the path integral.
The simulated experiments showed that momentum-based methods (i.e., NAG and Adam) could significantly accelerate the convergence on optimal control search.
We also applied the NAG-accelerated path integral method to MPC and PI-Net, 
demonstrating the improvement of control performance and the efficiency on PI-Net training for inverse optimal control.

The results of this work suggest several directions for future research.
First, although the path integral method derived in this paper is less constrained than the original methods, we did not analyze this in detail. 
This generalized property allows us to consider non-Gaussian settings and to design more expressive trajectory cost functions $S(\tau)$. 
We will further examine the effectiveness in order to broaden the applicability in practical systems.

Moreover, the causes of Adam's instability and AdaGrad's poor performance are not fully understood.
A suggestion from this paper is to utilize NAG because it is simple to apply and showed reliable acceleration for all the cases in our test.
However, further examination and understanding of these optimization methods will yield faster convergence.
In addition, the state-of-the-art optimization methods developed in the deep learning field, such as \cite{andrychowicz2016learning}, can be good solutions for further improvement.

Finally, constructing real-hardware demonstrations of MPC and inverse optimal control, utilizing our accelerated method, are on-going work.

\bibliography{icra}

\begin{thebibliography}{10}

\bibitem{kappen2005linear}
H.~J. Kappen, ``Linear theory for control of nonlinear stochastic systems,''
  {\em Phys. Rev. Lett.}, vol.~95, no.~20, 2005.

\bibitem{williams2016aggressive}
G.~Williams, P.~Drews, B.~Goldfain, J.~M. Rehg, and E.~A. Theodorou,
  ``Aggressive driving with model predictive path integral control,'' in {\em
  ICRA}, 2016.

\bibitem{williams2017model}
G.~Williams, A.~Aldrich, and E.~A. Theodorou, ``Model predictive path integral
  control: From theory to parallel computation,'' {\em J. Guid. Control Dyn.},
  2017.

\bibitem{Williams-ICRA-17}
G.~Williams, N.~Wagener, B.~Goldfain, P.~Drews, J.~Rehg, B.~Boots, and
  E.~Theodorou, ``Information theoretic {MPC} for model-based reinforcement
  learning.,'' in {\em ICRA}, 2017.

\bibitem{williams2017information}
G.~Williams, P.~Drews, B.~Goldfain, J.~M. Rehg, and E.~A. Theodorou,
  ``Information theoretic model predictive control: Theory and applications to
  autonomous driving,'' {\em arXiv:1707.02342}, 2017.

\bibitem{okada2017path}
M.~Okada, L.~Rigazio, and T.~Aoshima, ``Path integral networks: End-to-end
  differentiable optimal control,'' {\em arXiv:1706.09597}, 2017.

\bibitem{camacho2013model}
E.~F. Camacho and C.~B. Alba, {\em Model predictive control}.
\newblock Springer Science \& Business Media, 2013.

\bibitem{miyashita2017mirror}
M.~Miyashita, S.~Yano, and T.~Kondo, ``Mirror descent search and
  acceleration,'' {\em arXiv:1709.02535}, 2017.

\bibitem{theodorou2010generalized}
E.~Theodorou, J.~Buchli, and S.~Schaal, ``A generalized path integral control
  approach to reinforcement learning,'' {\em JMLR}, vol.~11, 2010.

\bibitem{bubeck2015convex}
S.~Bubeck {\em et~al.}, ``Convex optimization: Algorithms and complexity,''
  {\em Foundations and Trends{\textregistered} in Machine Learning}, 2015.

\bibitem{krichene2015accelerated}
W.~Krichene, A.~Bayen, and P.~L. Bartlett, ``Accelerated mirror descent in
  continuous and discrete time,'' in {\em NIPS}, 2015.

\bibitem{nesterov1983method}
Y.~Nesterov, ``A method of solving a convex programming problem with
  convergence rate {$O(1/k^2)$},'' {\em Soviet Mathematics Doklady}, 1983.

\bibitem{duchi2011adaptive}
J.~Duchi, E.~Hazan, and Y.~Singer, ``Adaptive subgradient methods for online
  learning and stochastic optimization,'' {\em JMLR}, 2011.

\bibitem{kingma2014adam}
D.~Kingma and J.~Ba, ``Adam: A method for stochastic optimization,'' {\em arXiv
  preprint arXiv:1412.6980}, 2014.

\bibitem{krichene2015efficient}
W.~Krichene, S.~Krichene, and A.~Bayen, ``Efficient bregman projections onto
  the simplex,'' in {\em CDC}, 2015.

\bibitem{todorov2005generalized}
E.~Todorov and W.~Li, ``A generalized iterative {LQG} method for
  locally-optimal feedback control of constrained nonlinear stochastic
  systems,'' in {\em ACC}, 2005.

\bibitem{wang1996approach}
H.~O. Wang, K.~Tanaka, and M.~F. Griffin, ``An approach to fuzzy control of
  nonlinear systems: Stability and design issues,'' {\em IEEE Trans. Fuzzy
  Syst.}, vol.~4, no.~1, 1996.

\bibitem{seguchi2003nonlinear}
H.~Seguchi and T.~Ohtsuka, ``Nonlinear receding horizon control of an
  underactuated hovercraft,'' {\em Int. J. Robust Nonlin}, 2003.

\bibitem{michael2010grasp}
N.~Michael, D.~Mellinger, Q.~Lindsey, and V.~Kumar, ``The {GRASP} multiple
  micro {UAV} testbed,'' {\em IEEE Robot. Autom. Mag.}, vol.~17, no.~3, 2010.

\bibitem{gonzales2016autonomous}
J.~Gonzales, F.~Zhang, K.~Li, and F.~Borrelli, ``Autonomous drifting with
  onboard sensors,'' in {\em International Symposium on Advanced Vehicle
  Control}, 2016.

\bibitem{andrychowicz2016learning}
M.~Andrychowicz, M.~Denil, S.~Gomez, M.~W. Hoffman, D.~Pfau, T.~Schaul, and
  N.~de~Freitas, ``Learning to learn by gradient descent by gradient descent,''
  in {\em NIPS}, 2016.

\end{thebibliography}
\bibliographystyle{ieeetr}

\begin{appendices}
\section{Derivation of the Original Iterative Path Integral Method} \label{app:it_derivation}
Let $p^{*}$ be the optimal probability density with $\mbm^{*}$.
We also define $p_{0}$ as the density of uncontrolled sequence $\mbm = \mathbf{0}$.
In \cite{Williams-ICRA-17}, the relation of these distributions are proved to be:
\begin{equation}
	p^{*} = \frac{p_{0}e^{-S_{\mbx}(\tau) / \lambda}}{\mathbb{E}_{p_{0}}\left[e^{-S_{\mbx}(\tau) / \lambda}\right]}. \label{eqn:optimal_dist}
\end{equation}
%
% One may notice the similarity between this equation and (\ref{eqn:update_law_p}).
Applying the similar approach with (\ref{eqn:update_law_immature}), (\ref{eqn:update_law_mu0}) results in the closed-form optimal solution: $\mbm^{*} = {\mathbb{E}_{p_{0}}\left[e^{-S_{\mbx}(\tau) / \lambda}\mbe\right]}/{\mathbb{E}_{p_{0}}\left[e^{-S_{\mbx}(\tau) / \lambda}\right]}$.
%
% \begin{equation}
% 	\mbm^{*} = \frac{\mathbb{E}_{p_{0}}\left[e^{-S(\tau) / \lambda}\mbe\right]}{\mathbb{E}_{p_{0}}\left[e^{-S(\tau) / \lambda}\right]}. \label{eqn:optimal_dist}
% \end{equation}
%
However this solution requires sampling of trajectories from the zero-mean distribution $p_{0}$.
Since this kind of sampling is inefficient requiring almost infinite samples,
Ref.~\cite{Williams-ICRA-17} alternatively employs iterative importance sampling from $p^{(j-1)}$:
\begin{equation}
	\begin{split}
	p^{(j)} &= \frac{p^{(j-1)}/p^{(j-1)}}{p^{(j-1)}/p^{(j-1)}}\cdot\frac{p_{0}e^{-S_{\mbx}(\tau) / \lambda}}{\mathbb{E}_{p_{0}}\left[e^{-S_{\mbx}(\tau) / \lambda}\right]} \\
    &= \frac{p^{(j-1)} \cdot \left[(p_{0}/p^{(j-1)})\cdot e^{-S_{\mbx}(\tau) / \lambda}\right]}{\mathbb{E}_{p^{(j - 1)}}\left[(p_{0}/p^{(j-1)}) \cdot e^{-S_{\mbx}(\tau) / \lambda}\right]}. \label{eqn:importance_sampling}
    \end{split}
\end{equation}
Considering that $p$ is Gaussian, we can represent the likelihood ratio $p_{0} / p^{(j-1)}$ as exponential form. Then integrating (\ref{eqn:importance_sampling}) yields the original iterative method (\ref{eqn:update_law}), (\ref{eqn:S_tilde}). We note that $R = \lambda \Sigma^{-1} / 2$ is supposed in the derivation.

\section{Cost functions} \label{app:cost}
This section summarizes the cost functions used in the experiment.
For all tasks, terminal cost was defined as same with running cost: $\phi(\mbx) = q(\mbx)$.

\paragraph{Pendulum Swing-up}
\begin{equation}
  \begin{split}
	q(\mbx) &= (1 + \cos\theta)^{2} + \dot{\theta}^{2}, \\
    R &= 5,
  \end{split}
\end{equation}
where $\theta$ is the angle of the pendulum.

\paragraph{Hovercraft Navigation}
\begin{equation}
  \begin{split}
    q(\mbx) &= h(d, w_{d}) + h(v, w_{v}) + h(\cos\theta_{d} - 1, w_{\theta}) \\
    &+ w_{F} \cdot (F_{1}^{2} + F_{2}^{2}), \\
    R &= O,
  \end{split}
\end{equation}
where, 
\begin{equation}
h(x, w) = \sqrt{x^{2} + w^{2}} - w,
\end{equation}
$d$ and $\theta_{d}$ are the distance and angular differences between current state and target state respectively, $v$ is the velocity, and $(w_{d}, w_{v}, w_{\theta},  w_{F}) = (10^{-6}, 10^{-2}, 1, 0.2)$.

\paragraph{Quadrotor Navigation}
\begin{equation}
  \begin{split}
    q(\mbx) &= h(\Delta x, w_{x}) + h(\Delta y, w_{y}) + h(\Delta z, w_{z})\\
    &+ w_{v} \cdot v^2 + w_{q} \cdot ||\Delta\mathbf{q}|| + w_{\boldsymbol\omega} \cdot ||{\boldsymbol\omega}|| + w_{\boldsymbol\Omega} \cdot ||{\boldsymbol\Omega}||, \\
    R &= O,
%     &+ h(\cos\theta_{d} - 1, w_{\theta}) + w_{F} \cdot (F_{1}^{2} + F_{2}^{2})
  \end{split}
\end{equation}
where $\Delta x$, $\Delta y$, $\Delta z$ are differences between current positions and target positions, $v$ is the velocity, $\Delta \mathbf{q}$ is the difference between current orientation and desired orientation, ${\boldsymbol \omega}$ is the angular velocity and $(w_{x,y,z}, w_{v}, w_{q}, w_{\boldsymbol\omega}, w_{\boldsymbol\Omega}) = (50, 10, 200, 10^{-3}, 10^{-6})$.

\paragraph{Car Navigation}
\begin{equation}
  \begin{split}
    q(\mbx) &=  100\left\{\left(\frac{x}{2}\right)^{2} + {y}^{2} - 1\right\}^{2} + (v_{x} - 1.25)^{2}, \\
    R &= O,
%     &+ h(\cos\theta_{d} - 1, w_{\theta}) + w_{F} \cdot (F_{1}^{2} + F_{2}^{2})
  \end{split}
\end{equation}
where $v_{x}$ is the forward velocity of the car.

\end{appendices}

\section*{Acknowledgment}
The authors would like to thank Shiro Yano and Megumi Miyashita for insightful discussion.

\end{document}